\documentclass[twocolumn,prb, amsmath,amssymb,aps,superscriptaddress,floatfix,nobibnotes]{revtex4-2}
\usepackage{CJK}
\usepackage{graphicx}
\usepackage{color}
\usepackage{braket}

\usepackage{xr-hyper}
\makeatletter
\newcommand*{\addFileDependency}[1]{
  \typeout{(#1)}
  \@addtofilelist{#1}
  \IfFileExists{#1}{}{\typeout{No file #1.}}
}
\makeatother

\newcommand*{\myexternaldocument}[1]{%
    \externaldocument{#1}%
    \addFileDependency{#1.tex}%
    \addFileDependency{#1.aux}%
}

\myexternaldocument{suppl}

\usepackage{dcolumn} 
\usepackage{bm}     
\usepackage{amsfonts}
\usepackage[bookmarks=false,colorlinks,citecolor=blue]{hyperref}
\hypersetup{colorlinks=true,linkcolor=blue,filecolor=blue,citecolor = blue, urlcolor=blue}\usepackage{amsmath}

\usepackage[version=4]{mhchem}
\usepackage{siunitx}

\newcommand{\bfl}{\begin{flushleft}}
\newcommand{\efl}{\end{flushleft}}
\begin{document}

\title{Synthesis and electronic properties of Nd$_{n+1}$Ni$_{n}$O$_{3n+1}$ Ruddlesden-Popper nickelate thin films}

\date{\today}
\author{Grace A. Pan}
\thanks{These authors contributed equally to this work.}
\affiliation{Department of Physics, Harvard University, Cambridge, MA, USA}
\author{Qi Song}
\thanks{These authors contributed equally to this work.}
\affiliation{Department of Physics, Harvard University, Cambridge, MA, USA}
\author{Dan Ferenc Segedin}
\affiliation{Department of Physics, Harvard University, Cambridge, MA, USA}
\author{Myung-Chul Jung}
\affiliation{Department of Physics, Arizona State University, Tempe, AZ, USA}
\author{Hesham El-Sherif}
\affiliation{The Rowland Institute, Harvard University, Cambridge, MA, USA}
\author{Erin E. Fleck}
\affiliation{School of Applied and Engineering Physics, Cornell University, Ithaca, NY, USA}
\author{Berit H. Goodge}
\affiliation{School of Applied and Engineering Physics, Cornell University, Ithaca, NY, USA}
\affiliation{Kavli Institute at Cornell for Nanoscale Science, Cornell University, Ithaca, NY, USA}
\author{Spencer Doyle}
\affiliation{Department of Physics, Harvard University, Cambridge, MA, USA}
\author{Denisse C\'{o}rdova Carrizales}
\affiliation{Department of Physics, Harvard University, Cambridge, MA, USA}
\author{Alpha T. N'Diaye}
\affiliation{Advanced Light Source, Lawrence Berkeley National Laboratory, Berkeley, CA, USA}
\author{Padraic Shafer}
\affiliation{Advanced Light Source, Lawrence Berkeley National Laboratory, Berkeley, CA, USA}
\author{Hanjong Paik}
\affiliation{Platform for the Accelerated Realization, Analysis and Discovery of Interface Materials (PARADIM), Cornell University, Ithaca, NY, USA}
\author{Lena F. Kourkoutis}
\affiliation{School of Applied and Engineering Physics, Cornell University, Ithaca, NY, USA}
\affiliation{Kavli Institute at Cornell for Nanoscale Science, Cornell University, Ithaca, NY, USA}
\author{Ismail El Baggari}
\affiliation{The Rowland Institute, Harvard University, Cambridge, MA, USA}
\author{Antia S. Botana}
\affiliation{Department of Physics, Arizona State University, Tempe, AZ, USA}
\author{Charles M. Brooks}
\affiliation{Department of Physics, Harvard University, Cambridge, MA, USA}
\author{Julia A. Mundy}
\thanks{mundy@fas.harvard.edu}
\affiliation{Department of Physics, Harvard University, Cambridge, MA, USA}

\begin{abstract}
The rare-earth nickelates possess a diverse set of collective phenomena including metal-to-insulator transitions, magnetic phase transitions, and, upon chemical reduction, superconductivity.  Here, we demonstrate epitaxial stabilization of layered nickelates in the Ruddlesden-Popper form, Nd$_{n+1}$Ni$_n$O$_{3n+1}$, using molecular beam epitaxy.  By optimizing the stoichiometry of the parent perovskite NdNiO$_3$, we can reproducibly synthesize the $n = 1 - 5$ member compounds.  X-ray absorption spectroscopy at the O $K$ and Ni $L$ edges indicate systematic changes in both the nickel-oxygen hybridization level and nominal nickel filling from 3$d^8$ to 3$d^7$ as we move across the series from $n = 1$ to $n = \infty$.  The $n = 3 - 5$ compounds exhibit weakly hysteretic metal-to-insulator transitions with transition temperatures that depress with increasing order toward NdNiO$_3$ ($n = \infty)$.  
\end{abstract}

\maketitle

\section{Introduction}

\begin{figure*}[t!]
    \centering
    \includegraphics[width =2\columnwidth]{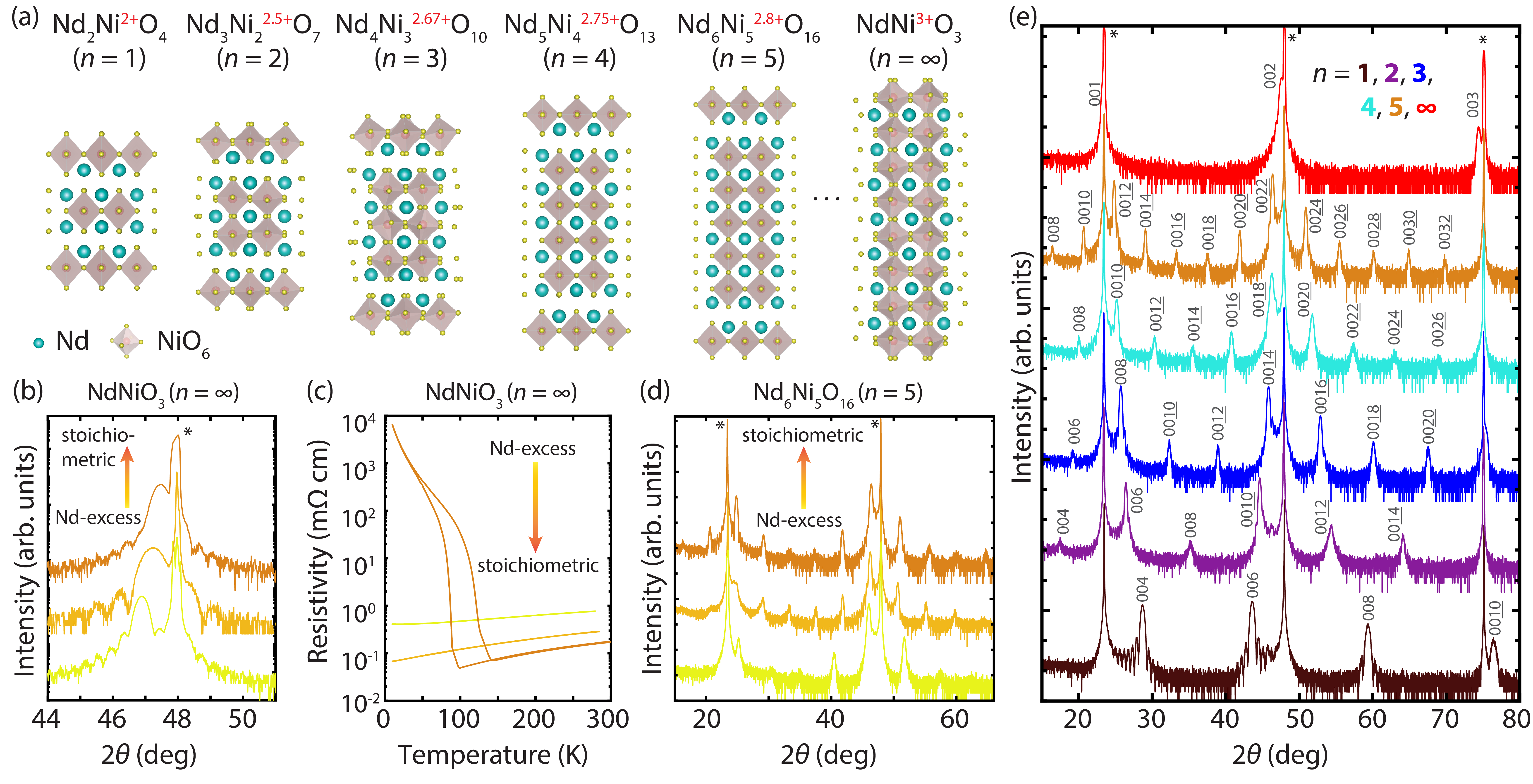}
    \caption{Structural description and characterization of the Ruddlesden-Popper  Nd$_{n+1}$Ni$_n$O$_{3n+1}$ compounds. (a) Crystal structures of the Nd$_{n+1}$Ni$_n$O$_{3n+1}$.  Nd, Ni, and O in turquoise, red, and yellow, respectively.  The $n = 4$ and $n = 5$ are drawn without octahedral rotations as there is no existing crystallographic data. (b) X-ray diffraction (XRD) spectra and (c) electrical resistivity characterization of two separate parent perovskite NdNiO$_3$ calibrations that were used to synthesize the two $n = 5$ compounds in (d).  Although both NdNiO$_3$ films are phase pure and show sub-nanometer surface roughness, using the calibration from an off-stoichiometric calibration forms a higher-order Ruddlesden-Popper of reduced crystallinity.  (e) XRD spectra of all the Nd$_{n+1}$Ni$_n$O$_{3n+1}$ ($n = 1-5$) synthesized with optimized NdNiO$_3$ calibrations.  Asterisks denote substrate peaks. 
    }
    \label{fig:structural}
\end{figure*}

The perovskite rare earth nickelates $R$NiO$_3$ ($R$ = La, Pr, Nd...) are strongly correlated materials with a rich, tunable phase diagram that includes features such as metal-to-insulator transitions and non-collinear antiferromagnetism \cite{Torrance1992,catalano2018rare}.
The Ruddlesden-Popper nickelates, represented by the chemical formula $R_{n+1}$Ni$_n$O$_{3n+1}$, are the layered analogues of the perovskite rare-earth nickelates.  Also expressed as ($R$NiO$_3$)$_n$($R$O),  these compounds comprise $n$ layers of the traditional perovskite motif $R$NiO$_3$ separated by rock salt spacer layers $R$-O as shown in Fig.\ \ref{fig:structural}(a).  The $R$-O layer tunes the dimensionality of the system from the three-dimensional $R$NiO$_3$ ($n = \infty$) to the quasi two-dimensional $R_2$NiO$_4$ ($n$ = 1).  The phenomenology of the $R_2$NiO$_4$ family of compounds is distinct from its perovskite counterparts: for example, while the perovskite LaNiO$_3$ is a strongly correlated metal \cite{Torrance1992,sreedhar1992electronic,liu2020observation},  single-layer La$_2$NiO$_4$ displays insulating charge and spin stripes, with a spatial modulation tunable with doping \cite{rodriguez1991neutron,emery}.  Accordingly, the evolution of the charge and spin structure with perovskite layer thickness (or `order') $n$ in the Ruddlesden-Popper nickelates has been of sustained interest, especially as many of these ordered phases precede superconductivity in the cuprates \cite{Keimer2015}.   Indeed, the Ruddlesden-Popper nickelates have long been viewed as close cousins of the cuprates \cite{GREENBLATT1997174, lacorre1991synthesis,nat_phys} and it was recently found that the $n = 5, \infty$ compounds exhibit superconductivity when chemically reduced into the square-planar form \cite{Li2019,pan2022superconductivity}. 

Despite longstanding interest, synthesis of the Ruddlesden-Popper nickelates remains a key challenge.  Early studies on the $n = 2, 3$ member compounds in powder form identified analogies to the superconducting cuprates including a putatively similar band structure and possible charge density wave transitions \cite{zhang1995synthesis,seo1996electronic,GREENBLATT1997174,poltavets2006oxygen}.  More recently, advances in the preparation of large single-crystal specimens \cite{nat_phys,zhang2020high} have enabled direct study of these features which were originally proposed from powder specimen behavior \cite{li2017fermiology,zhang2020intertwined}.
Nevertheless, the higher-order compounds ($n \geq 4$), which are of particular interest as the $n = 5$ becomes superconducting when reduced \cite{pan2022superconductivity}, remain inaccessible using powder or single crystal synthetic methods \cite{zhang2020high}.  Instead, these compounds require atomic layering control.  In addition to the synthesis of the superconducting $n = 5$ square-planar compound with molecular beam epitaxy (MBE), the stabilization of the La$_{n+1}$Ni$_{n}$O$_{3n+1}$\cite{Li2020_RP_phases} and  Nd$_{n+1}$Ni$_{n}$O$_{3n+1}$\cite{sun2021electronic} series for $n = 1 - 5$ also using MBE has recently been reported.  Atomically-precise thin film deposition thus allows access to the  $R_{n+1}$Ni$_n$O$_{3n+1}$ series beyond $n = 3$. 

In this paper, we describe a generalizable synthetic strategy for thin film Ruddlesden-Popper nickelates Nd$_{n+1}$Ni$_n$O$_{3n+1}$ ($n$ = 1 – 5), using MBE.  By first optimizing the crystalline growth of the perovskite NdNiO$_3$ on LaAlO$_3$ substrates, we can reproducibly stabilize the higher-order layered Ruddlesden-Popper nickelates. From x-ray absorption spectra (XAS), we observe that tuning the layering $n$ of the Ruddlesden-Popper nickelates changes the nickel electronic filling and relative transition-metal-oxygen hybridization levels, as would be expected from electron counting rules.  Furthermore, there are systematic changes in the x-ray linear dichroism (XLD) which we attribute to energy level splittings in the $e_g$ orbitals.  In the electronic transport, we observe in the $n \geq 2 $ compounds metal-to-insulator transitions with features that interpolate between those of the parent perovskite NdNiO$_3$ and bulk crystal specimens of Nd$_{n+1}$Ni$_n$O$_{3n+1}$ ($n$ = 3) \cite{GREENBLATT1997174}.

\section{Experimental Methods}

\subsection{Synthesis via MBE}

We use ozone-assisted MBE to synthesize the Ruddlesden-Popper nickelates in the thin film form.   The evolution of the nickel valence from Ni$^{2+}$ in the Nd$_2$NiO$_4$ ($n = 1$) compound to Ni$^{3+}$ in the NdNiO$_3$ compound necessitates varied synthetic conditions \cite{zhang2020high}—as hot as 1000\textdegree C for  Nd$_2$NiO$_4$ and as cold as 550\textdegree C for NdNiO$_3$.  We synthesize the Nd$_2$NiO$_4$ ($n = 1$) compound using the simultaneous evaporation of both elements (hereafter, `codeposition') at high ($\sim$900-1000 \textdegree C) temperatures with fluxes estimated using a quartz crystal microbalance with applied tooling factors.  We also note that it is possible to synthesize the $n$ = 2, 3 and perovskite layer compounds using the codeposition method by tuning only the relative flux ratios, temperature and oxidation pressure, similar to what has been done for single crystals \cite{zhang2020high}.   On the other hand, the $n \geq 4$ compounds, which have not been synthesized as bulk crystals, also cannot be synthesized with codeposition.  These compounds therefore require the precise sequential deposition of the neodymium and nickel sources, wherein a single monolayer of each element is deposited at a time, to achieve the Ruddlesden-Popper layering.

In principle, MBE enables one to precisely control the monolayer deposition times and hence synthesize a generic Ruddlesden-Popper compound of arbitrary order $n$.  There are key differences however between the Ruddlesden-Popper nickelates and other Ruddlesden-Popper systems such as the more commonly studied Sr$_{n+1}$Ti$_n$O$_{3n+1}$ which make the direct translation of other synthetic calibration techniques onto the nickelates difficult.  In Sr$_{n+1}$Ti$_n$O$_{3n+1}$ compounds, deviations from perfect stoichiometry by as much as 5+\% often result in imperfect superlattices with rock salt intergrowths or missing rock salt layers but without substantial phase segregation \cite{dawley2020defect,brooks2009growth}.  Errors in monolayer dosing times can then be quantitatively estimated and adjusted for purely based on x-ray diffraction spectra of the superlattice phase \cite{barone2021improved}. The composition, or $A$:$B$ cation ratio in an $AB$O$_3$ compound, can also be perfected using quantitative analysis of the beat frequencies and oscillation lineshapes in the reflection high energy electron diffraction (RHEED) intensities \cite{haeni2000rheed,brooks2009growth}.  

In contrast, a principal challenge in nickelate thin film synthesis is the stabilization of the high oxidation Ni$^{3+}$ state.  Insufficient oxidizing conditions can promote the phase segregation of Ni$^{2+}$ compounds such as polycrystalline NiO \cite{lacorre1991synthesis,li2021impact}.  The presence of NiO however, can result not only from insufficient oxidation, but also from errors in composition or monolayer dosing, and most frequently from a combination of all three.  This propensity to phase segregate makes assessing quantitative changes to monolayer dosing challenging.  Moreover, it is difficult to use RHEED oscillations to precisely adjust for composition errors: once a secondary phase such as NiO forms, RHEED oscillation intensities fade and do not recover, whereas oscillations can persist for much longer in SrTiO$_3$-based compounds \cite{haeni2000rheed,brooks2009growth}.  

To address these challenges we have developed an alternative calibration method that reliably facilitates the synthesis of $n \geq 2$ Ruddlesden-Popper Nd$_{n+1}$Ni$_n$O$_{3n+1}$ compounds.  This method optimizes the synthesis of the parent perovskite NdNiO$_3$: by iteratively improving both its stoichiometric composition and monolayer dose estimates, we can obtain precise shuttering times to synthesize the higher order Ruddlesden-Popper compounds with atomistic precision.  We start by adjusting the temperature of the neodymium and nickel effusion cells to read $\sim 1 \times 10^{13}$ atoms cm$^{-1}$s$^{-1}$ on a quartz crystal microbalance.  We then synthesize the binary oxides Nd$_2$O$_3$ on Y:ZrO$_2$ (111) and NiO on MgO (100) and use x-ray reflectivity (XRR) to estimate the film thicknesses and thus the actual effusion cell fluxes \cite{sun2022canonical}.  These estimates give us approximate values of the neodymium and nickel fluxes, which we then fine tune with the synthesis of NdNiO$_3$.

From the rough flux estimates provided by the binary oxide compounds, we perform the shuttered growth of NdNiO$_3$.  In our initial calibration scheme, we intentionally deposit less than one full monolayer of both neodymium and nickel.  The accidental deposition of more than one full monolayer can exacerbate the formation of polycrystalline NiO phases, which would be observed in the RHEED (Fig.\ S4 of the Supplemental Material \cite{suppl}), even if the elements are supplied in a one-to-one composition ratio.  Hence in this ``underdosed" regime, we can focus on tuning the composition ratio of neodymium and nickel.  Similar to other oxide perovskites \cite{brooks2009growth}, the parent compound NdNiO$_3$ is capable of accommodating some $A$-site (neodymium) excess into the lattice, while still forming a single-phase film, as ascertained by RHEED and atomic force microscopy (AFM).  This $A$-site excess is manifest as an expansion of the out-of-plane lattice constant as well as in a broadening or even suppression of the metal-to-insulator transition \cite{breckenfeld2014effects,preziosi2017reproducibility}.   By tracking the evolution of the lattice constants for a range of Nd:Ni ratios, we have found that an optimally stoichiometric NdNiO$_3$ film on LaAlO$_3$ possesses a lattice constant of $\sim$3.82 \AA{} corresponding to a (002) film peak that nearly coincides with that of the substrate LaAlO$_3$ \cite{breckenfeld2014effects,preziosi2017reproducibility}.  The metal-to-insulator transition in these stoichiometric films spans as much as over five orders of magnitude; films with excess neodymium have a higher room temperature resistivity and may possess a entirely suppressed metal-to-insulator transition.  We note that it is also possible to form a nickel-rich NdNiO$_3$ phase \cite{li2021impact}, though we have found that nickel excess will more frequently appear as a secondary NiO phase \textit{in situ} in the RHEED pattern (Fig.\ S4 in the Supplemental Material \cite{suppl}).  After fine-tuning this shuttered stoichiometric calibration in the underdosed regime, we then correct for the monolayer dosing using thickness estimates from XRR fits to the NdNiO$_3$ films.  Finally, we apply the monolayer deposition times from the calibrated NdNiO$_3$ to synthesize the Ruddlesden-Popper compounds. We synthesize the Nd$_{n+1}$Ni$_n$O$_{3n+1}$ ($n \geq 2$) compounds at a substrate temperature of 630 – 690\textdegree C as verified by an optical pyrometer and using distilled ozone in chamber pressures of 1.2$\times 10^{-6}$ – 2.0$\times 10^{-6}$ Torr.

Figures \ref{fig:structural}(b)-\ref{fig:structural}(d) illustrates this approach using three different calibrations from NdNiO$_3$ films.  The stoichiometric film possesses a lattice constant of 3.82(7) \AA; in contrast, the $A$-site-rich films have expanded lattice constants of 3.84(5) and 3.87(3) \AA{}, corresponding to neodymium in excess of $\sim$3\% and $\sim$10\%, respectively, estimated from our neodymium flux calibrations (Fig.\ \ref{fig:structural}(b)).  Importantly, we note that this $A$-site excess is not always manifest in surface-sensitive characterization such as RHEED or AFM (Fig.\ S4 in the Supplemental Material \cite{suppl}) images, which show that the $A$-site excess films are ostensibly high quality.  Hence, we confirm the superior quality of the stoichiometric film using resistivity measurements which present a sharp metal-to-insulator transition in Fig. \ref{fig:structural}(c).  Finally, using the three different shuttered calibrations from these three NdNiO$_3$ films, we synthesize the Nd$_6$Ni$_5$O$_{16}$ ($n$ = 5) compound.  As shown in the XRD spectra in Fig.\ \ref{fig:structural}(d),  the Nd$_6$Ni$_5$O$_{16}$ synthesized from the two $A$-site excess films display predominant Ruddlesden-Popper layering but with weaker superlattice ordering.  The Nd$_6$Ni$_5$O$_{16}$ film synthesized from the least stoichiometric NdNiO$_3$ calibration possesses superlattice peaks approaching those of an $n = 4$ compound, and neither film exhibits sharp lower order peaks.  In contrast, the Nd$_6$Ni$_5$O$_{16}$ film synthesized from the stoichiometric NdNiO$_3$ calibration displays coherent $n = 5$ ordering with the full emergence of the lower order (00\underline{10}) and (00\underline{12}) superlattice peaks, which are sensitive to the long-range order of the film.  This indicates the sensitivity of the Ruddlesden-Popper nickelates to composition, requiring a calibration procedure that enables fine tuning.

\begin{figure*}
    \centering
    \includegraphics[width =2\columnwidth]{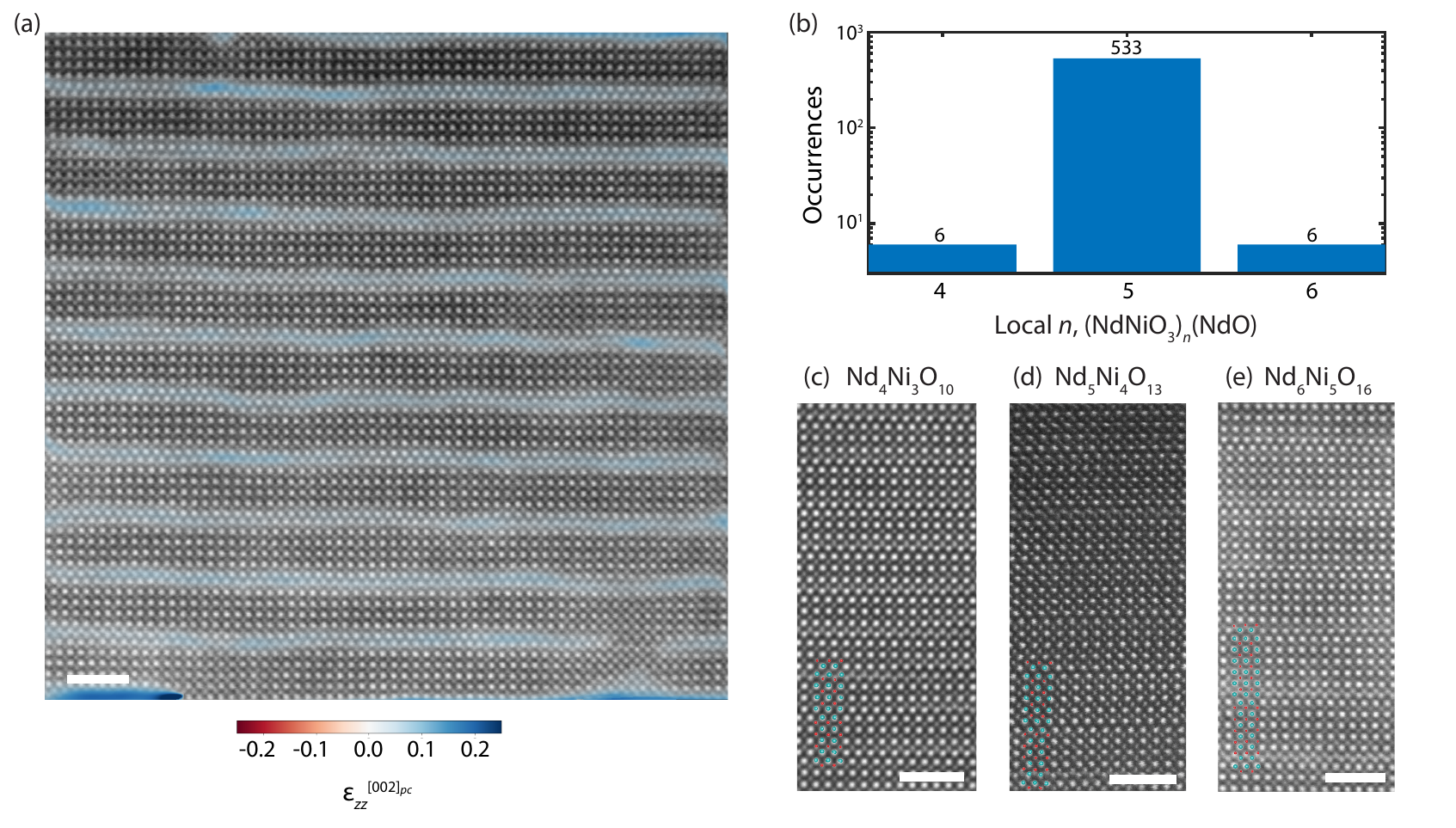}
    \caption{High-angle annular dark-field scanning transmission electron microscopy (HAADF-STEM) images of Ruddlesden-Popper Nd$_{n+1}$Ni$_n$O$_{3n+1}$ compounds. (a) Large field of view of the Nd$_{6}$Ni$_5$O$_{16}$ ($n = 5$) compound with an overlaid strain map to highlight the Ruddlesden-Popper layers.  Scale bar, 2 nm.  (b) Bar graph showing the number of occurrences of each of the Ruddlesden-Popper layers over the same field of view, plotted on a log scale.  (c)-(e) High-resolution images of the (c) $n = 3$, (d) $n = 4$, and (e) $n = 5$ compounds, illustrating the precise placement of the Nd-O rock salt double layers every $n$ perovskite layers.  Neodymium and nickel atoms are represented by turquoise and red, respectively.  Scale bars, 2 nm.
    }
    \label{fig:tem}
\end{figure*}

\subsection{Structural characterization}

Thin film XRD spectra were taken on Malvern Panalytical Empyrean diffractometer using Cu K$\alpha_1$ ($\lambda = 1.5406$ \AA) radiation.  Reciprocal space maps (RSMs) were taken on the same instrument using a PIXcel3D 2D pixel area detector.  Cross-section scanning transmission electron microscopy (STEM) specimens were prepared using an FEI Helios focused-ion beam with a gallium ion source.  The final thinning step was performed using 2 keV gallium ions.  Atomic-resolution STEM imaging was performed using a Thermo Fisher Scientific Themis Z G3 transmission electron microscope operated at 200 keV with probe convergence angle of $\sim$19.6 mrad and an annular dark-field (ADF) collection angle range of $\sim$78--200 mrad.  Additional STEM imaging was performed on an aberration-corrected JEOL ARM 200F transmission electron microscope at 200 keV with a probe convergence angle of $\sim$22 mrad and an ADF collection angle range of $\sim$68--220 mrad.  The displayed images were obtained from the average of 20 cross-correlated frames each acquired with 500 nS dwell time.

To map the local distortions of specific lattice fringes, we performed a phase lock-in analysis on the high-angle ADF STEM (HAADF-STEM) images. By taking the gradient of the local phase, we can generate a map of local strain along the $c$ direction \cite{elbaggari2018nature,goodge2022disentangling, fleck2022semi}.  The strain map highlights Ruddlesden-Popper rock salt layers, which appear as regions of local tensile strain beacuse the spacing between two adjacent Nd-O layers ($\sim$2.7 \AA{}) is greater than the spacing between the adjacent Nd-O and Ni-O layers ($\sim$1.9 \AA{}). Analyzing the strain profiles along the $c$-direction can thus be used to quantify distances between the rock salt space layers. This allows us to determine the locations and occurrences of local $n$ layerings within a film, as demonstrated in Fig.\ \ref{fig:tem}.  Further details on the phase lock-in analysis may be found in the Supplemental Material \cite{suppl}.

\begin{table*}[t]
\begin{tabular}{ |p{.9cm}||p{4.5cm}|p{3.0cm}|p{3.9cm}|p{3.2cm}|  }
 \hline
      Order & Hypothetical valence (filling)  & Measured valence & $c$-axis lattice constant (\AA) & Unit-cell volume (\AA$^3$) \\
 \hline
 $n$ = 1 & 2+ ($d^8$) & 2+ (nominal)  & 12.450 $\pm$ 0.002 & 361 $\pm$ 0.06\\
 $n$ = 2 & 2.5+ ($d^{7.5}$) & 2.5(7)+ & 20.20 $\pm$ 0.05 & 580 $\pm$ 1.5 \\
 $n$ = 3 & 2.67+ ($d^{7.33}$) &  2.6(7)+ & 27.728 $\pm$ 0.015 & 797 $\pm$ 0.43 \\
 $n$ = 4 & 2.75+ ($d^{7.25}$) &  2.7(9)+ & 35.35 $\pm$ 0.07 &  1016 $\pm$ 2\\
 $n$ = 5 & 2.8+  ($d^{7.2}$) & 2.8(4)+ & 43.12 $\pm$ 0.04 & 1239 $\pm$ 1.2\\

 \hline
\end{tabular}

\caption{\label{tab:doping} Nominal nickel valences (hypothetical) and structural parameters (experimental) for  Nd$_{n+1}$Ni$_{n}$O$_{3n+1}$ epitaxially stabilized on LaAlO$_3$. A 2+ valence state is assumed for the $n = 1$ compound.}
\end{table*}

\subsection{X-ray absorption spectroscopy}

XAS were measured at the Advanced Light Source, Lawrence Berkeley National Lab, at Beamlines 4.0.2 and 6.3.1 in the total electron yield mode at 300 K.  At Beamline 4.0.2 the spectra were acquired at 20\textdegree{} grazing incidence with linear horizontally ($I_z$) or vertically ($I_x$) polarized photons.  At Beamline 6.3.1, the spectra were acquired exclusively with linear horizontally polarized incident photons but with the sample either normal ($I_x$) or at 30\textdegree{} grazing ($I_z$) to the beam.  A geometric correction factor was applied for the grazing incident signal \cite{wu2004orbital}.  As energy calibrations can differ slightly between beamline endstations, we use Nd $M_{4,5}$ features to align all Ni $L_{2,3}$- and O $K$-edge spectra.  
All spectra are normalized to the incident x-ray flux, as monitored via the absorption by a semitransparent gold mesh upstream of the nickelate films.  The spectra are further scaled so that the intensity is unity at energies slightly below the absorption edge, and normalized across the entire edge.  Each spectrum presented is polarization-averaged [$\frac{1}{2}(I_x+I_z)$] and represents the mean of 8-16 polarization-averaged scans.  All x-ray linear dichroism (XLD) signals are given by $I_z - I_x$ ($\mathbf{E} || \mathbf{c} - \mathbf{E} || \mathbf{a,b}$), the difference between spectra with photons polarized predominantly out-of-plane and in-plane with respect to the sample surface.  The XLD spectra are normalized by the polarization-averaged integral $A = 2(I_x+I_z)/3$ \cite{wu2013strain}.

\subsection{Electrical transport}

Electrical resistivities were determined using devices in van der Pauw or Hall bar geometries.  Contacts comprising Cr (5 nm)/Au (100 nm) were deposited using an electron-beam evaporator and patterned with shadow masks.  Hall bar channels were defined with a diamond scribe.  Transport measurements down to 1.8 K were conducted in a Quantum Design Physical Property Measurement System equipped with a 9 T magnet using AC lock-in techniques at $\sim$ 15 Hz.  Temperature-dependent Hall coefficients were calculated from linear fits of anti-symmetrized field sweeps up to 9 T, all taken upon warming.

\section{Results and Discussion}

\subsection{Thin film growth}

\begin{figure*}[t]
    \centering
    \includegraphics[width =2\columnwidth]{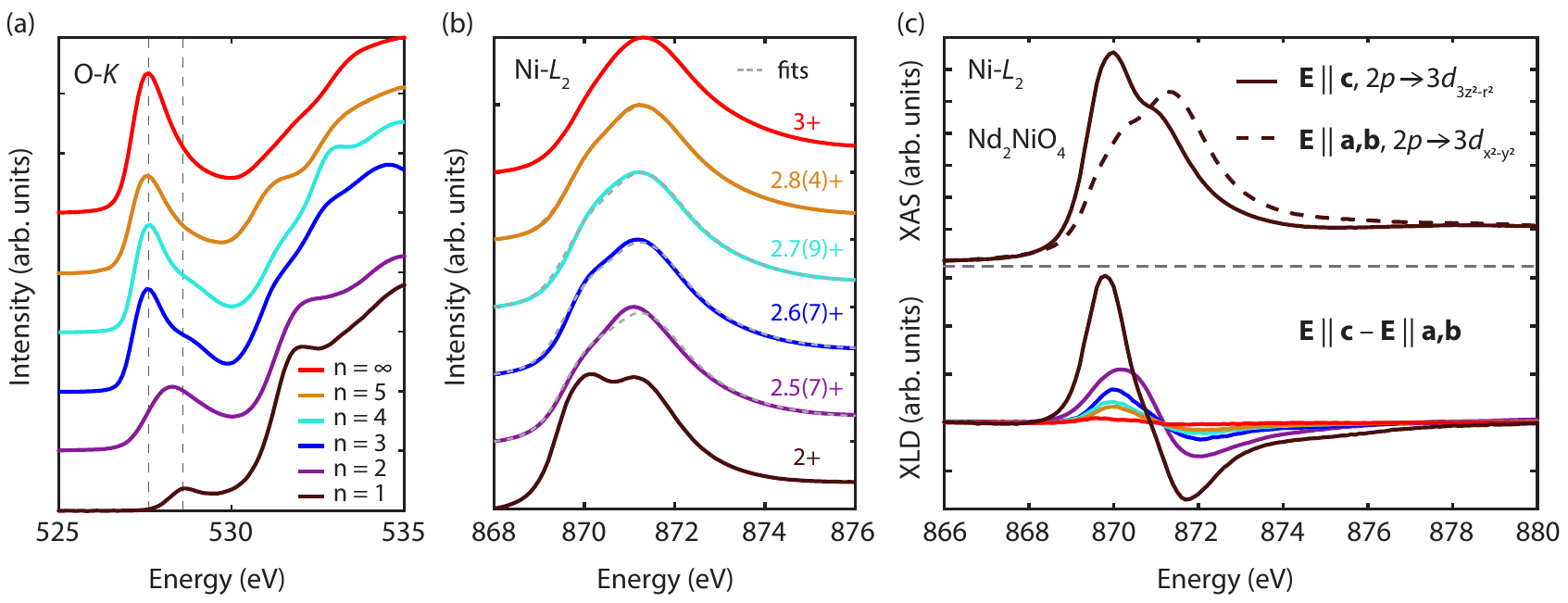}
    \caption{X-ray absorption spectra (XAS) of the Ruddlesden-Popper Nd$_{n+1}$Ni$_n$O$_{3n+1}$ compounds. (a) Oxygen $K$-edge prepeak features.  Dashed lines represent the spectral weights of transitions to the nominal 3$d^7$ (left) and 3$d^8$ (right) states. (b) Spectra across the Ni $L_2$-edge.  Data are in solid curves.  Fits to the data with non-negative linear least squares fittings using the $n = 1$ and $n = \infty$ spectra assigned as references are shown as dashed light grey curves.  Estimated valences from the fits are shown on the side. (c) X-ray linear dichroic (XLD) signals (bottom) at the Ni $L_2$-edge, represented as $\mathbf{E} || \mathbf{c} - \mathbf{E} || \mathbf{a,b}$.  An example of the polarization-dependent XAS signals used to determine the XLD from the $n = 1$ compound is shown at the top.}
    \label{fig:xas}
\end{figure*}

Using the growth calibration techniques described above, we present the XRD spectra of the fully epitaxially stabilized nickelate Ruddlesden-Poppers ($n =$ 1 – 5) on LaAlO$_3$ in Fig.\ \ref{fig:structural}(e).  The XRD spectra display all the allowed even-numbered (00$l$) superlattice peaks.  We extract the out-of-plane lattice constants from Nelson-Riley fits to the 2$\theta$-$\theta$ spectra \cite{nelsonriley}, tabulated in table \ref{tab:doping}.  

To determine the in-plane lattice constants, we perform RSMs on each of the films (see Fig.\ S2 in the Supplemental Material \cite{suppl}).  RSM scans indicate that the $n = 2 - 5, \infty$ films are epitaxially strained to the LaAlO$_3$ substrate ($a = 3.79$ \AA), which provides $\varepsilon \approx -0.4\%$ strain to the bulk NdNiO$_3$ ($n = \infty$) and as much as $\varepsilon \approx -0.9\%$ strain to the bulk Nd$_4$Ni$_3$O$_{10}$ ($n = 3$).  However, our single-layer Nd$_2$NiO$_4$ ($n = 1$) film, which experiences a similar strain level of $\varepsilon \approx -0.9\%$, appears to have partially relaxed to an average $a \approx 3.81 $ \AA. This is still under the bulk lattice constant of $a_{\mathrm{pc}} = 3.825$ \AA.  This relaxation is likely order-dependent; for example, in the single-layer Ruddlesden-Popper ruthenate Sr$_2$RuO$_4$, immediate relaxation has been reported on substrates providing $\varepsilon \gtrsim -0.9\%$ strain \cite{burganov2016strain}.  Indeed, while our Nd$_2$NiO$_4$ ($n = 1$) film exhibits partial relaxation as early as 17 nm, we could stabilize the Nd$_4$Ni$_3$O$_{10}$ ($n = 3$) to around $\sim$55 nm before we observed partial relaxation (see Fig.\ S3 of the Supplemental Material \cite{suppl}).  The films presented in Fig. \ref{fig:structural}(e) were all synthesized to have 60--75 nickel layers, resulting in thicknesses of 29.5, 27.7, 26.5, 32.4, and 22.9 nm for the $n = 2 - 5, \infty$ compounds, respectively; meanwhile, the $n = 1$ compound presented is 17 nm thick.  With these in-plane lattice constants, we report the unit cell volumes in Table \ref{tab:doping}.  Compared with the bulk lattice constants of Nd$_4$Ni$_3$O$_{10}$ \cite{zhang2020high,li2021contrasting,zhang1995synthesis,li2020metal,rout2020structural} we observe a slight decrease in the overall unit cell volume by $\sim$0.7\% with the compressive epitaxial strain imposed by the LaAlO$_3$ substrate, as has also been observed in NdNiO$_3$ \cite{kim2020strain}.  A more detailed discussion on the determination of lattice parameters and errors can be found in the Supplemental Material \cite{suppl}.

HAADF-STEM images taken across the series confirms the coherent layering at the microscopic level, as shown in Fig.\ \ref{fig:tem}.  In Fig.\ \ref{fig:tem}(a), we observe a long-range coherent ordering of the horizontal Ruddlesden-Popper structure up to the total thickness of a 32-nm-thick $n = 5$ film, with no obvious vertical intergrowths observed within the largest fields-of-view of our imaging.  In addition, atomic contrast provided by HAADF-STEM highlights the placement of the Nd-O rock salt layers every $n$ unit cells of the perovskite motif, as exemplified in the close up images in Figs.\ \ref{fig:tem}(c)-\ref{fig:tem}(e) for the $n \geq 3$ compounds.  We do however observe defect regions of reduced atomic contrast between the neodymium and nickel sites (Fig.\ S7 in the Supplemental Material \cite{suppl}).  We ascribe these to half-unit-cell offsets between regions of the film, likely caused by the occasional stacking fault, as opposed to the possibility of cation intermixing.  Since STEM is a measurement in projection, these offsets would create regions where scattering intensities are averaged over both neodymium and nickel sites in the atomic columns.

In other Ruddlesden-Popper systems, areas of locally varying $n$ are frequently observed \cite{lee2013exploiting,lee2014dynamic,nie2014atomically,barone2021improved,nair2018demystifying}.  To quantify the distribution of the Ruddlesden-Popper layerings, we employ phase lock-in analysis using the (001)$_{\mathrm{pc}}$ and (101)$_{\mathrm{pc}}$ peaks of the NdNiO$_3$ perovskite structure \cite{elbaggari2018nature,goodge2022disentangling, fleck2022semi}.  Figure \ref{fig:tem}(e) displays the strain map produced from the phase lock-in analysis, we which we have superimposed on the bare STEM image.  The map highlights Ruddlesden-Popper rock salt layers, which appear as regions of local tensile strain.  We find that $n = 5$ is indeed the dominant Ruddlesden-Popper layering for the nominal Nd$_6$Ni$_5$O$_{12}$ film presented in Figs.\ \ref{fig:structural}(e) and \ref{fig:tem}(b), though there are occasional occurrences of $n = 4$ and 6 Ruddlesden-Popper layerings.  A phase lock-in analysis is also sensitive to defect structures or anywhere there is a deviation from the periodicity corresponding to the Fourier peak, as illustrated in Fig.\ S7 in the Supplemental Material \cite{suppl}.  Additional details on the quantification of Ruddlesden-Popper layerings may be found in the Supplemental Material \cite{suppl}.

\subsection{Electronic structure}

To measure electronic structure changes across the Ruddlesden-Popper series, we use XAS.  We start with the O-$K$ edges, which can provide a probe of the nickel valence.  In particular, we focus on the \textit{prepeak}, or first peak along the oxygen edge, which arises from the covalent mixing of the oxygen 2$p$ with the nickel 3$d$ states \cite{deGroot1989oxygen}.  Due to varying contributions from the LaAlO$_3$ substrate $\gtrsim 530$ eV, we do not address the higher-energy fine structure features, whereas the lower-energy pre-peak region is free from background features.  The full O $K$-edge spectra can be found in Fig.\ S8 in the Supplemental Material \cite{suppl}.  The intensity of this prepeak across members of a homologous transition metal series should scale with the number of unoccupied 3$d$ states (increasing valence state) as per a 1$s$ to hybridized 2$p$-3$d$ (or 3$d\underline{L}^n$ type transition) \cite{deGroot1989oxygen,suntivich2014estimating}.  Here then, Nd$_2$NiO$_4$ with a nominal valence of Ni$^{2+}$ should have the weakest intensity O $K$ prepeak with a minor shift to higher energies representing transitions into states of the type $\alpha \ket{3d^8}+\beta \ket{3d^9\underline{L}}$ \cite{kuiper1991unoccupied,pellegrin19961s,kuiper1998polarization}.  Meanwhile NdNiO$_3$, with a valence of Ni$^{3+}$ and a ground state dominated by highly covalent $3d^8\underline{L}^n$ states \cite{mizokawa2000spin,bisogni2016ground} (nominally $3d^7$) should have the strongest intensity prepeak, with all intermediate Ruddlesden-Poppers interpolating in-between.  Indeed we observe this general trend in pre-peak intensity across our Nd$_{n+1}$Ni$_n$O$_{3n+1}$ samples, as shown in Fig.\ \ref{fig:xas}(a).  We quantify both the integrated pre-peak intensities, which scale with the nickel-oxygen hybridization levels, and the maximum of the prepeak signal, which reflects the relative number of unoccupied states, as shown in Fig.\ S8 in the Supplemental Material \cite{suppl}.  In both quantities, and particularly in the estimates of the integrated pre-peak intensities, the differences between the higher order ($n = 3-5$) compounds are within error.  This indicates that relative nickel-oxygen hybridization levels do not appreciably change between Ruddlesden-Popper compounds of increasing high order, unlike in their reduced square-planar counterparts \cite{pan2022superconductivity}.  Nevertheless, the general trend of increased pre-peak intensity from $n$ = 1 to $\infty$ points to an overall depletion of the nominal nickel 3$d$ states and increasing covalency as we move down the Ruddlesden-Popper series from NdNiO$_3$ ($n = \infty$) to Nd$_2$NiO$_4$ ($n$ = 1).

To further assess the filling of the nickel 3$d$ states, we can decompose this pre-peak region into the two distinct features demarcated by the dotted grey lines in Fig. \ref{fig:xas}(a).  We assign these features to represent transitions associated with the hybridized (nominal) $3d^7$ and $3d^8$ states \cite{pellegrin19961s,hu2000hole}.  Accordingly, the mixed-valence Ruddlesden-Poppers ($n = 2 - 5$) possess spectral weight from both these features with more of the lower energy (higher valence) feature as we increase the dimensionality of the system.  This effect has been previously observed in single-layer $R_{2-x}$Sr$_x$NiO$_4$ compounds \cite{kuiper1991unoccupied,pellegrin19961s}, where hole-doping away from Ni$^{2+}$ toward Ni$^{3+}$ shifts the total spectral weight of the O $K$ prepeak toward lower incident energy in XAS.  With the recent observation of superconductivity in a layered nickelate \cite{pan2022superconductivity} and the identification of inequivalent transition-metal planes in a five-layer cuprate \cite{kunisada2020observation}, the layer-resolved distribution of the nickel valence is of increasing interest though conflicting reports exist in bulk crystal studies \cite{zhang2020high, li2020metal}.  This observation that the mixed-valence Ruddlesden-Popper compounds possess both Ni$^{2+}$ and Ni$^{3+}$ features in the O-$K$ edge pre-peak region, in contrast to previous work \cite{di2021spectroscopic}, provides a future means to use spatially resolved spectroscopic techniques such as electron-energy loss spectroscopy to identify any electronic modulation.

The Ni $L_{2,3}$-edges, which we present in Fig. \ref{fig:xas}(b), are comparatively more difficult to interpret due to core-hole effects \cite{fink19852p}, and are exquisitely sensitive to variations in the strain state, cation composition, and oxygen content \cite{kuiper1998polarization,freeland2016evolution,kim2020strain, tung2017polarity}.  The strong covalency effects in the oxidized nickelates complicate the quantitative extraction of valence as can be performed in other perovskite oxides such as the titanates or manganites \cite{deGroot1989oxygen,kourkoutis2010atomic,mundy2014visualizing}.  Nevertheless, we can employ a self-consistent method to characterize the systematic evolution of the nickel valence across the series, as has been similarly performed using x-ray photoemission spectroscopy \cite{sun2021electronic}.  We define our spectra from the Nd$_2$NiO$_4$ ($n$ = 1) and NdNiO$_3$ ($n = \infty$) end-member compounds, which have been well-characterized in the bulk \cite{piamonteze2005spin,hu2000hole}, as proxies for the Ni$^{2+}$ and Ni$^{3+}$ reference spectra, respectively, within the layered nickelate series.  We then use these reference spectra in a non-negative linear least squares fitting to extract the nominal nickel valences as tabulated in Table \ref{tab:doping}.  As the octahedral coordination environments and transition metal--oxygen hybridization should be reasonably similar across the Ruddlesden-Popper nickelates, we expect this approach to be more reflective of relative valence changes than using reference spectra from generic Ni$^{2+}$ compounds such as NiO.  The valences interpolate nearly monotonically between 2+ and 3+ as we move from $n = 1$ to $\infty$ and are close to the expected values from simple electron counting rules.  (Note that we employ this method on just the higher energy Ni $L_{2}$-edges, which are free of the La $M_{4}$ background from the LaAlO$_3$ substrate that runs into the Ni $L_3$-edge as seen in Fig.\ S8 in the Supplemental Material \cite{suppl}.)  We attribute deviations from the ideal fractional valences to minor variations in cation stoichiometry and oxygen content which are challenging to quantify precisely in nickelate thin films.  Nonetheless, this general procedure reveals the gradation of the nickel filling from formal 3$d^8$ to 3$d^7$ across the order of the Ruddlesden-Popper compounds, further indicating that the layering may be harnessed to control the electronic filling of the rare-earth nickelates.

The Ni $L_2$ spectra presented above are polarization averaged; we now decompose them to compare the x-ray linear dichroic (XLD) signals across the series.  At the Ni $L$-edge, we expect from dipole selection rules that only 2$p \rightarrow 3d_{3z^2-r^2}$ transitions are allowed for $\mathbf{E} || \mathbf{c}$ ($I_z$) and primarily 2$p \rightarrow 3d_{x^2-y^2}$ transitions are allowed for $\mathbf{E} || \mathbf{a,b}$ ($I_x$) \cite{pellegrin19961s}.  Hence, assuming Nd$_2$NiO$_4$ ($n = 1$) to be in a high-spin state with two half-filled $e_g$ orbitals we expect a large XLD signal arising from a splitting of the $3d_{3z^2-r^2}$ and $3d_{x^2-y^2}$ orbitals generated by a strong tetragonal distortion \cite{pellegrin19961s,kuiper1998polarization}.  In contrast, the dichroism in the NdNiO$_3$ ($n = \infty$) film is negligible: with a single half-filled $e_g$ orbital, no dichroism is expected without substantial strain or heterostructure engineering that breaks the double degeneracy of the $e_g$ orbitals \cite{chaloupka2008orbital,disa2015orbital,wu2013strain,disa2015research}. Thus the XLD of the mixed-valent $n = 2-5$ compounds should decrease going from the single-layer Nd$_2$NiO$_4$ to the perovskite NdNiO$_3$.  This is indeed the trend observed experimentally in Fig. \ref{fig:xas}(c). 
Compressive strain may also contribute to a slight enhancement of this dichroic response.  It has been previously shown in nickelate compounds that in straightening out the NiO$_6$ octahedra in the $c$ axis and promoting a tetragonal distortion, compressive strain acts to lower the energy of the $3d_{3z^2-r^2}$ orbital and shifts the $\mathbf{E} || \mathbf{c}$ spectra to lower energy \cite{freeland2011orbital,tung2013connecting,disa2015research}.  As we move from the bulk lattice constants of $c$ = 3.807 \AA{} in NdNiO$_3$ ($n = \infty$) \cite{lacorre1991synthesis} to $c$ = 3.827 \AA{} in Nd$_2$NiO$_4$ ($n = 1$), the compressive strain increases from $\varepsilon\sim -0.45\%$ to  $-1\%$.  Hence, our observed trend of an increased XLD signal, with the $3d_{3z^2-r^2}$ ($\mathbf{E} || \mathbf{c}$) peak sitting at lower energy than the $3d_{x^2-y^2}$ ($\mathbf{E} || \mathbf{a,b}$), as we decrease the order $n$ of the Nd$_{n+1}$Ni$_n$O$_{3n+1}$ compounds may also reflect the additional effect of compressive epitaxial strain.

\subsection{Electrical transport}

\begin{figure}
    \centering
    \includegraphics[width = 1\columnwidth]{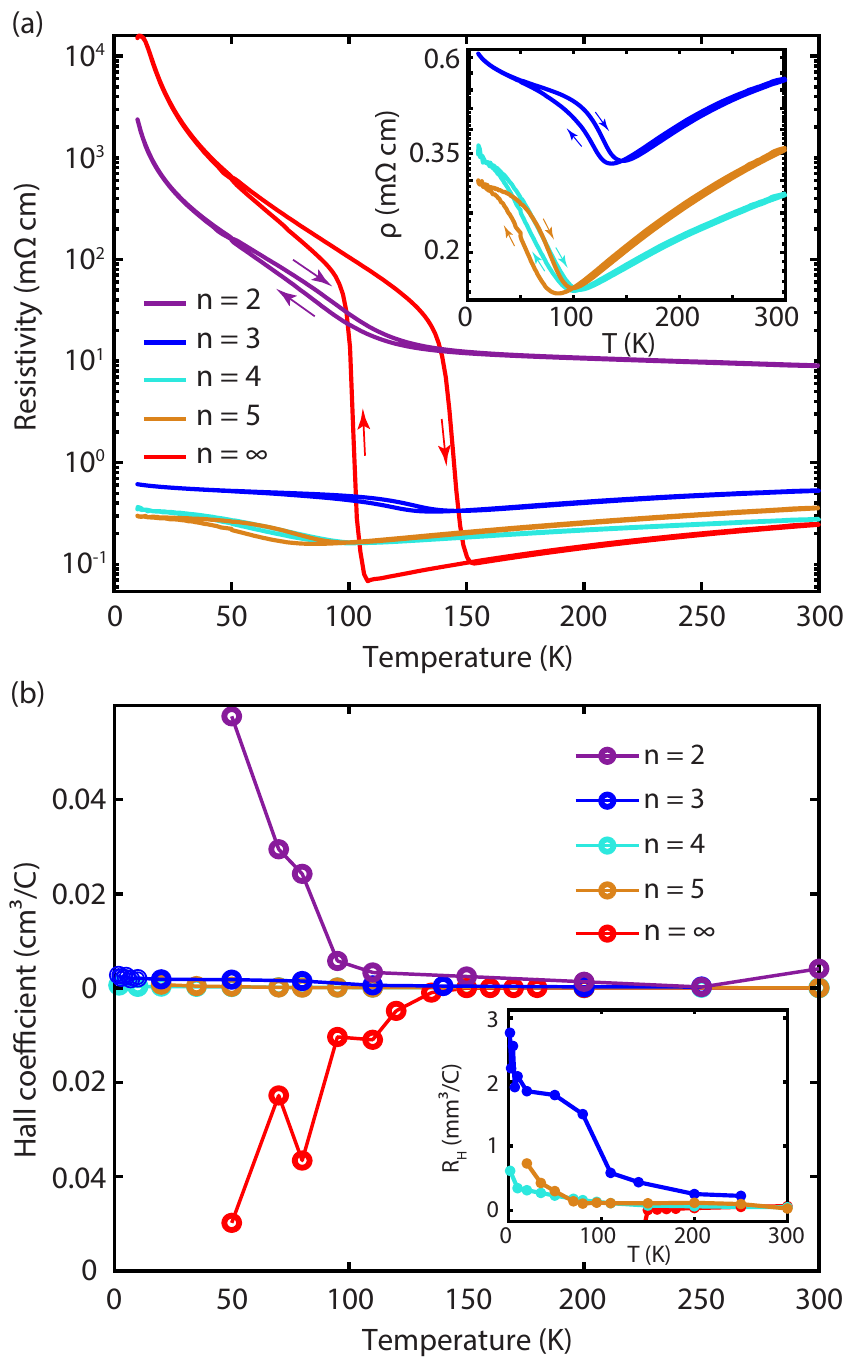}
    \caption{Electrical characterization of the Nd$_{n+1}$Ni$_n$O$_{3n+1}$ thin films.  (a) Temperature-dependent resistivity of the Nd$_{n+1}$Ni$_n$O$_{3n+1}$ for $n = 2-5, \infty$.  Arrows indicate the direction of temperature sweeps.  The $n = 1$ compound is insulating as in the bulk.  Inset is a zoom-in on the $n = 3-5$ compounds. (b)  Hall coefficients ($R_\mathrm{H}$) for the $n = 2 - 5, \infty$ compounds.  All the Ruddlesden-Popper compounds possess a sudden increase in $R_\mathrm{H}$ when cooled through the metal-to-insulator transitions. The $R_\mathrm{H}$ behavior for the NdNiO$_3$ ($n = \infty$) exhibits a sign change from positive to negative when cooled into the insulating state. Inset is a zoom-in on the $n = 3-5$ compounds.}
    \label{fig:transport}
\end{figure}

Having demonstrated the structural and electronic modulations of the Ruddlesden-Popper nickelates as a function of layering order $n$, we now present electrical resistivity measurements for the Nd$_{n+1}$Ni$_n$O$_{3n+1}$ for $n = 2 – 5$, $\infty$ in Fig. \ref{fig:transport}.  The $n = 1$ compound is insulating as in the bulk, likely due to the presence of a charge-stripe-ordered state \cite{emery}.  The $n = 2$ compound exhibits a near metal-to-insulating transition with minor hysteresis, though we have also observed fully insulating behavior in samples of slightly diminished quality, similar to observations in bulk La$_3$Ni$_2$O$_7$ \cite{zhang1994synthesis}.  The $n \geq 3$ compounds exhibit a metal-to-insulator transition with a hysteresis reminiscent of the perovskite NdNiO$_3$ albeit with a far weaker transition to the insulating state.  The transition temperatures span between $\sim$87 and 150 K and are generally suppressed with the increasing order of the system although there is slight variation sample to sample.  Intriguingly, both the metal-to-insulator transition temperature and magnitude of the resistivity jump through the transition decrease with the increasing order of the system.  This is contrary to an expectation that the resistivities of the Nd$_{n+1}$Ni$_n$O$_{3n+1}$, in moving to higher order $n$, would begin to approximate those of the parent NdNiO$_3$ $n = \infty$. 

The bulk $R_4$Ni$_3$O$_{10}$ compounds possess a second-order metal-to-metal transition at $\sim$ 150 K emblematic of the formation of a charge density wave (CDW) instability \cite{li2020metal,GREENBLATT1997174,li2021contrasting,rout2020structural}. Interestingly, we do not observe this behavior in our Nd$_4$Ni$_3$O$_{10}$ films stabilized with compressive epitaxial strain, though we have previously found this metal-to-metal transition in thin film Nd$_4$Ni$_3$O$_{10}$ on NdGaO$_3$ substrates which provide tensile strain \cite{pan2022superconductivity}.   Nevertheless, our compressively strained Nd$_4$Ni$_3$O$_{10}$ films may still possess a CDW transition that is now manifest as a metal-to-insulator not metal-to-metal transition.  The temperature-dependent Hall coefficients exhibit a sharp jump at the transition temperature as shown in Fig. \ref{fig:transport}(b).  This is consistent with bulk measurements reflecting a loss of carriers due a Fermi surface reconstruction and CDW instability in Nd$_4$Ni$_3$O$_{10}$ \cite{li2021contrasting}.  In contrast, the opening of a bond disproportionated gap in the NdNiO$_3$ compounds is reflected in a change in the sign of Hall carriers \cite{hauser2013temperature}.  Thus it is likely that the epitaxially stabilized Ruddlesden-Popper compounds still possess a CDW transition as in single crystal Nd$_4$Ni$_3$O$_{10}$ but that compressive epitaxial strain modifies this transition.  The precise role of strain on modulating these charge states in these Ruddlesden-Popper compounds would require further investigation. 
We note that the weak hysteresis and metal-to-insulator transitions -- characteristics which qualitatively resemble the resistivity behavior of NdNiO$_3$ -- appear intrinsic to thin film Nd$_{n+1}$Ni$_n$O$_{3n+1}$ ($n > 3$) stabilized on LaAlO$_3$, as they are reproducible across all high-quality films (see Fig. S5 in the Supplemental Material \cite{suppl}).  These features are likely not due to the presence of trace NdNiO$_3$, as we do not observe large areas with missing rock salt layers in STEM images, as shown in Fig.\ \ref{fig:tem}.
Finally, the evolution of the temperature-dependent Hall coefficients in Fig.\ \ref{fig:transport}(b) suggest that as we increase the order of the system, the low-temperature Hall coefficients decrease and thus, the bands crossing the Fermi level begin to approximate those in the parent NdNiO$_3$ \cite{hauser2013temperature}.  This points to a potential crossover from the incommensurate CDW in the $n = 3$ to the fully charge disproportionated state in the $n = \infty$ within the $3 < n < \infty$ range.

\section{Conclusions}

In summary, we have synthesized epitaxial thin films of the neodymium-based Ruddlesden-Popper nickelates, Nd$_{n+1}$Ni$_n$O$_{3n+1}$ with MBE.  Our synthetic strategy carefully optimizes the deposition of the perovskite NdNiO$_3$ to enable the precise layer-by-layer growth of the Ruddlesden-Popper films, and can potentially be generalized to any Ruddlesden-Popper system.  XAS at the O-$K$ and Ni-$L$ edges demonstrate a consistent depletion of the formal nickel 3$d^8$ states as we move across the series from $n = 1$ to $\infty$.  There is also a concomitant increase in the oxygen character as would be expected from increasing valence \cite{zaanen1985band}.   XLD suggests that the splitting between the in-plane and out-of-plane orbitals in the $e_g$ manifold increases, consistent with an increased tetragonal distortion toward $n = 1$.  In the electronic transport the $n \geq 3$ compounds exhibit a weakly hysteretic metal-to-insulator transition though with resistivity changes small compared with the parent NdNiO$_3$.  This resistivity behavior differs from the non-hysteretic, second-order metal-to-metal transitions in bulk $R_{n+1}$Ni$_n$O$_{3n+1}$ which are reflective of a CDW instability.  The temperature-dependent Hall coefficients however suggest that such a CDW-like transition does occur, but that this transition begins to cross over to the charge-ordered transition in NdNiO$_3$ ($n = \infty$) as $n$ increases.  This underscores the apparent sensitivity of these layered nickelates to both epitaxial strain and dimensionality.

Our work invites further exploration of the electronic and magnetic phases in thin film Ruddlesden-Popper systems, and proximate to the superconductivity observed in our reduced Nd$_{6}$Ni$_5$O$_{12}$ compound \cite{pan2022superconductivity}.  For example, bulk single-crystal specimens of Nd$_4$Ni$_3$O$_{10}$ exhibit intertwined charge and spin density waves \cite{zhang2020intertwined}.  Investigating the coupling of charge and spin order in the layered nickelates would thus be an exciting future pursuit, facilitated by the successful synthesis of thin-film Ruddlesden-Popper nickelates.  Finally, synthesis of Nd$_{n+1}$Ni$_n$O$_{3n+1}$ thin films provides a platform to search for additional superconductivity in the square-planar Nd$_{n+1}$Ni$_n$O$_{2n+2}$ compounds.

\vspace{5mm}

\section*{Acknowledgments}
\noindent We thank H. Hijazi at the Rutgers University Laboratory of Surface Modification for assistance in Rutherford backscattering spectrometry. Research is primarily supported by the US Department of Energy (DOE), Office of Basic Energy Sciences, Division of Materials Sciences and Engineering, under Award No. DE-SC0021925.  Materials growth was supported by PARADIM under National Science Foundation (NSF) Cooperative Agreement No. DMR-2039380.  Electron microscopy was primarily carried out through the use of MIT.nano facilities at the Massachusetts Institute of Technology.  Additional electron microscopy and all nanofabrication work was performed at Harvard University's Center for Nanoscale Systems, a member of the National Nanotechnology Coordinated Infrastructure Network, supported by the NSF under Grant No. 2025158.  In this paper, we also used resources of the Advanced Light Source, a DOE Office of Science User Facility under Contract No. DE-AC02-05CH11231.  G.A.P. acknowledges support from the Paul \& Daisy Soros Fellowship for New Americans and from  NSF Graduate Research Fellowship Grant No. DGE-1745303. Q.S. and D.C.C. were supported by the Science and Technology Center for Integrated Quantum Materials, NSF Grant No. DMR-1231319. M.-C.J and A.S.B. acknowledge NSF Grant No. DMR-2045826 and the ASU Research Computing Center for high-performance computing resources.  H.E.S. and I.E. were supported by the Rowland Institute at Harvard.  E.E.F., B.H.G., and L.F.K. were supported by the NSF (PARADIM) under Cooperative Agreement No. DMR-2039380.  J.A.M acknowledges support from the Packard Foundation and the Gordon and Betty Moore Foundation’s EPiQS Initiative, Grant No. GBMF6760. 

G.A.P. and Q.S. contributed equally to this work.  

\bibliography{bib.bib}

\end{document}



\title{Supplemental Material for: Synthesis and electronic properties of Nd$_{n+1}$Ni$_{n}$O$_{3n+1}$ Ruddlesden-Popper nickelate thin films}

\date{\today}
\author{Grace A. Pan}
\thanks{These authors contributed equally to this work.}
\affiliation{Department of Physics, Harvard University, Cambridge, MA, USA}
\author{Qi Song}
\thanks{These authors contributed equally to this work.}
\affiliation{Department of Physics, Harvard University, Cambridge, MA, USA}
\author{Dan Ferenc Segedin}
\affiliation{Department of Physics, Harvard University, Cambridge, MA, USA}
\author{Myung-Chul Jung}
\affiliation{Department of Physics, Arizona State University, Tempe, AZ, USA}
\author{Hesham El-Sherif}
\affiliation{The Rowland Institute, Harvard University, Cambridge, MA, USA}
\author{Erin E. Fleck}
\affiliation{School of Applied and Engineering Sciences, Cornell University, Ithaca, NY, USA}
\author{Berit H. Goodge}
\affiliation{School of Applied and Engineering Sciences, Cornell University, Ithaca, NY, USA}
\affiliation{Kavli Institute at Cornell for Nanoscale Science, Cornell University, Ithaca, NY, USA}
\affiliation{Department of Physics, Arizona State University, Tempe, AZ, USA}
\author{Spencer Doyle}
\affiliation{Department of Physics, Harvard University, Cambridge, MA, USA}
\author{Denisse C\'{o}rdova Carrizales}
\affiliation{Department of Physics, Harvard University, Cambridge, MA, USA}
\author{Alpha T. N'Diaye}
\affiliation{Advanced Light Source, Lawrence Berkeley National Laboratory, Berkeley, CA, USA}
\author{Padraic Shafer}
\affiliation{Advanced Light Source, Lawrence Berkeley National Laboratory, Berkeley, CA, USA}
\author{Hanjong Paik}
\affiliation{Platform for the Accelerated Realization, Analysis and Discovery of Interface Materials (PARADIM),Cornell University, Ithaca, NY, USA}
\author{Lena F. Kourkoutis}
\affiliation{School of Applied and Engineering Sciences, Cornell University, Ithaca, NY, USA}
\affiliation{Kavli Institute at Cornell for Nanoscale Science, Cornell University, Ithaca, NY, USA}
\author{Ismail El Baggari}
\affiliation{The Rowland Institute, Harvard University, Cambridge, MA, USA}
\author{Antia S. Botana}
\affiliation{Department of Physics, Arizona State University, Tempe, AZ, USA}
\author{Charles M. Brooks}
\affiliation{Department of Physics, Harvard University, Cambridge, MA, USA}
\author{Julia A. Mundy}
\thanks{mundy@fas.harvard.edu}
\affiliation{Department of Physics, Harvard University, Cambridge, MA, USA}
\maketitle
\newpage

\section*{Determination of structural parameters}

\begin{figure}[!h]
    \centering
    \includegraphics[width = \columnwidth]{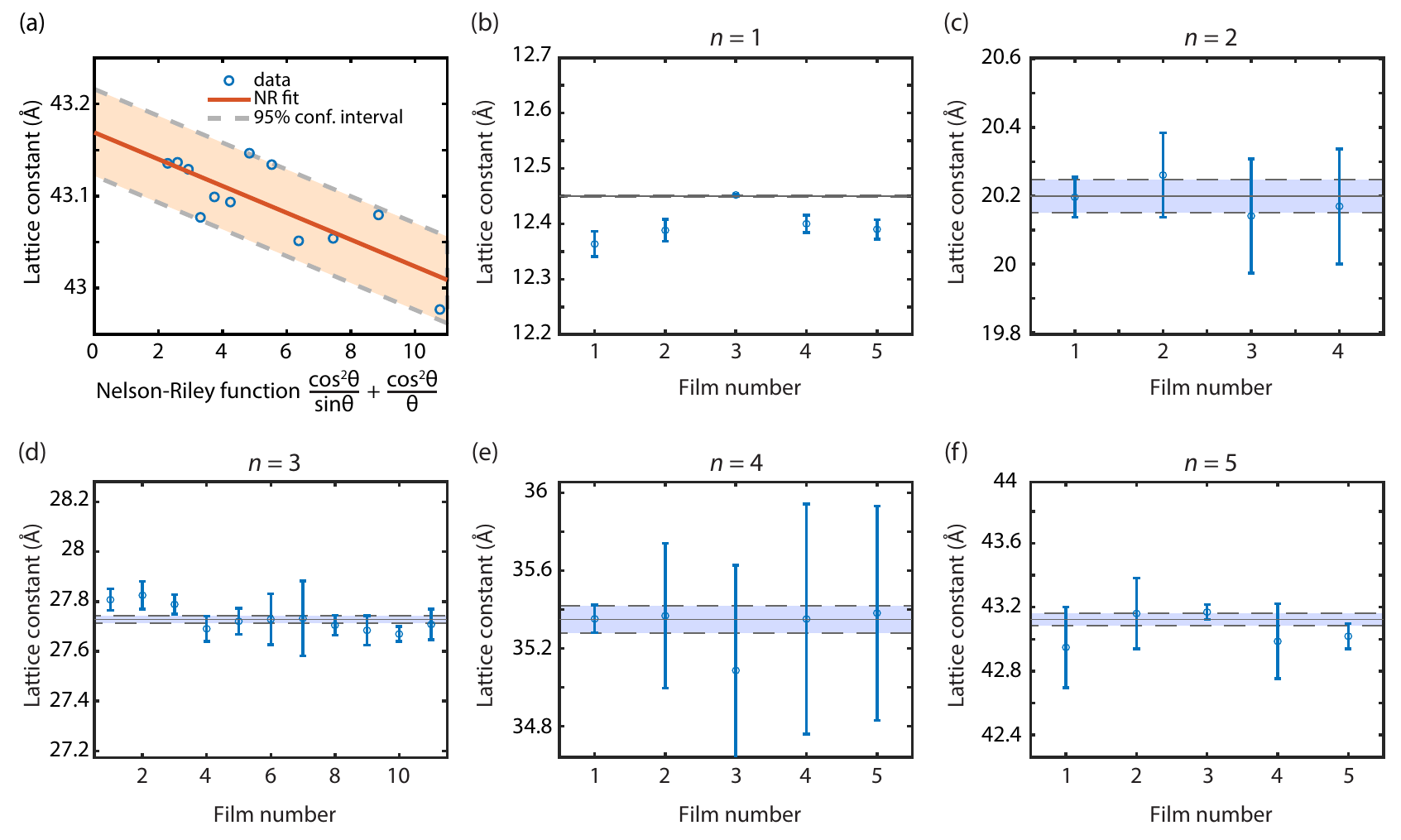}
    \caption{(a) Sample Nelson-Riley fit of an $n = 5$ film used in the determination of the out-of-plane lattice parameter.  The points plotted represent even-numbered superlattice reflections (00\underline{10}) to (00\underline{32}), from right to left.  The dashed lines represent the linear regression shifted to the bounds of a 95\% confidence interval estimating the y-intercept, i.e. the out-of-plane lattice constant. (b)-(e) Out-of-plane lattice constants estimated from individual higher-quality films for $n = 1-5$.  Error bars for each film are taken from the confidence interval bounds of the y-intercept in (a). The solid grey lines represent the best estimate of the lattice parameter as determined by a weighted mean; the dashed grey lines bound the uncertainties.  The film numbers are a counting index; films are plotted randomly.}
    \label{fig:NRfit}
\end{figure}

To accurately report on the bulk structural properties of the Ruddlesden-Popper thin films, we apply Nelson-Riley fitting to the x-ray diffraction spectra. \cite{nelsonriley}  The Nelson-Riley procedure is an empirical method of deriving unit cell lattice parameters by using all the superlattice reflections of a crystal and weighting each one by the Nelson-Riley function, $\frac{cos^2\theta}{sin\theta} + \frac{cos^2\theta}{\theta}$.  The nominal lattice parameters associated with the individual superlattice reflections are plotted against the Nelson-Riley function; the intercept of a linear regression then yields an estimate of the out-of-plane lattice constant.  An example of a Nelson-Riley fit performed on an $n = 5$ film is shown in Fig.\ \ref{fig:NRfit}, with a 95\% confidence interval around the estimation of the intercept, or out-of-plane lattice constant.

This procedure yields a more precise estimation of the lattice constant than taking an arithmetic mean of the lattice constants derived from individual superlattice reflections alone.  The Nelson-Riley function more strongly weights lower-angle superlattice reflections, which are sensitive to longer-range order and therefore more sensitive to structural quality. Thus in our lattice parameter estimates, we include only films that possess a fully formed (00\underline{2$n$+2}) reflection and clear indications of the (00\underline{2$n$}) peak, such as those shown in Fig.\ 1 and further exemplified in Fig.\ \ref{fig:variablefilms} in the next section. However, we exclude peaks with index lower than (00\underline{2$n$}) because the NR function effectively diverges at low angle ($2\theta \lesssim 15^{\circ}$) such that the error in estimating low angle peak positions dominates over the fitting error.  Fig.\ \ref{fig:NRfit}(b)-(e) show the spread in lattice constants for films of order $n = 1 - 5$. Error bars per film $\sigma_i$ are given as the 95\% confidence interval of the intercept taken from the Nelson-Riley linear regression procedure.  The final lattice constants per order $n$ we report in the table 1 of main text are determined by a weighted average $\sum x_i w_i/\sum w_i$. This quantity represents a weighting of each lattice constant $x_i$ per film by a weight $w_i = \frac{1}{\sigma_i}$.  The total uncertainty per order $n$ is then given by 1/$\sqrt{\sum w_i}$, represented by the blue-shaded box in Fig.\ \ref{fig:NRfit}(b)-(e).

\begin{figure}[!h]
    \centering
    \includegraphics[width = \columnwidth]{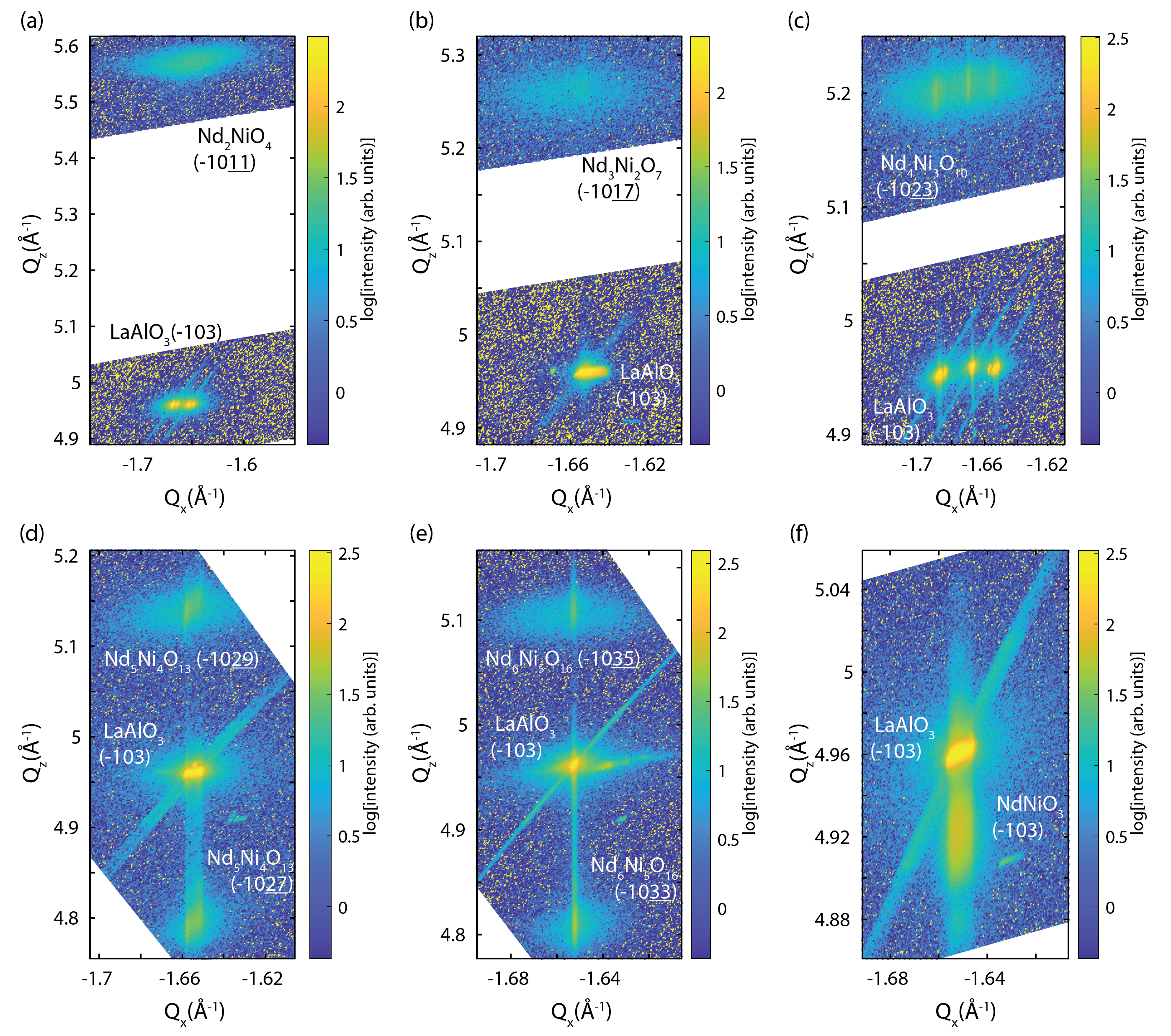}
    \caption{Reciprocal space maps of the Nd$_{n+1}$Ni$_n$O$_{3n+1}$, $n = 1-5$ (a)-(e) and NdNiO$_3$ (f) demonstrating that all films are epitaxially strained to the substrate LaAlO$_3$ ($c$ = 3.79 \AA, $\varepsilon \approx -(0.4$-$1)\%$).  All peaks are labelled in the pseudocubic notation.  The split peaks, most prominent in the $n = 1$ (a) and $n = 3$ (c) films, arise from the rhombohedral twinning of the LaAlO$_3$ substrate.  This twinning can vary dramatically substrate to substrate.}
    \label{fig:rsm}
\end{figure}

\begin{figure}[!h]
    \centering
    \includegraphics[width = \columnwidth]{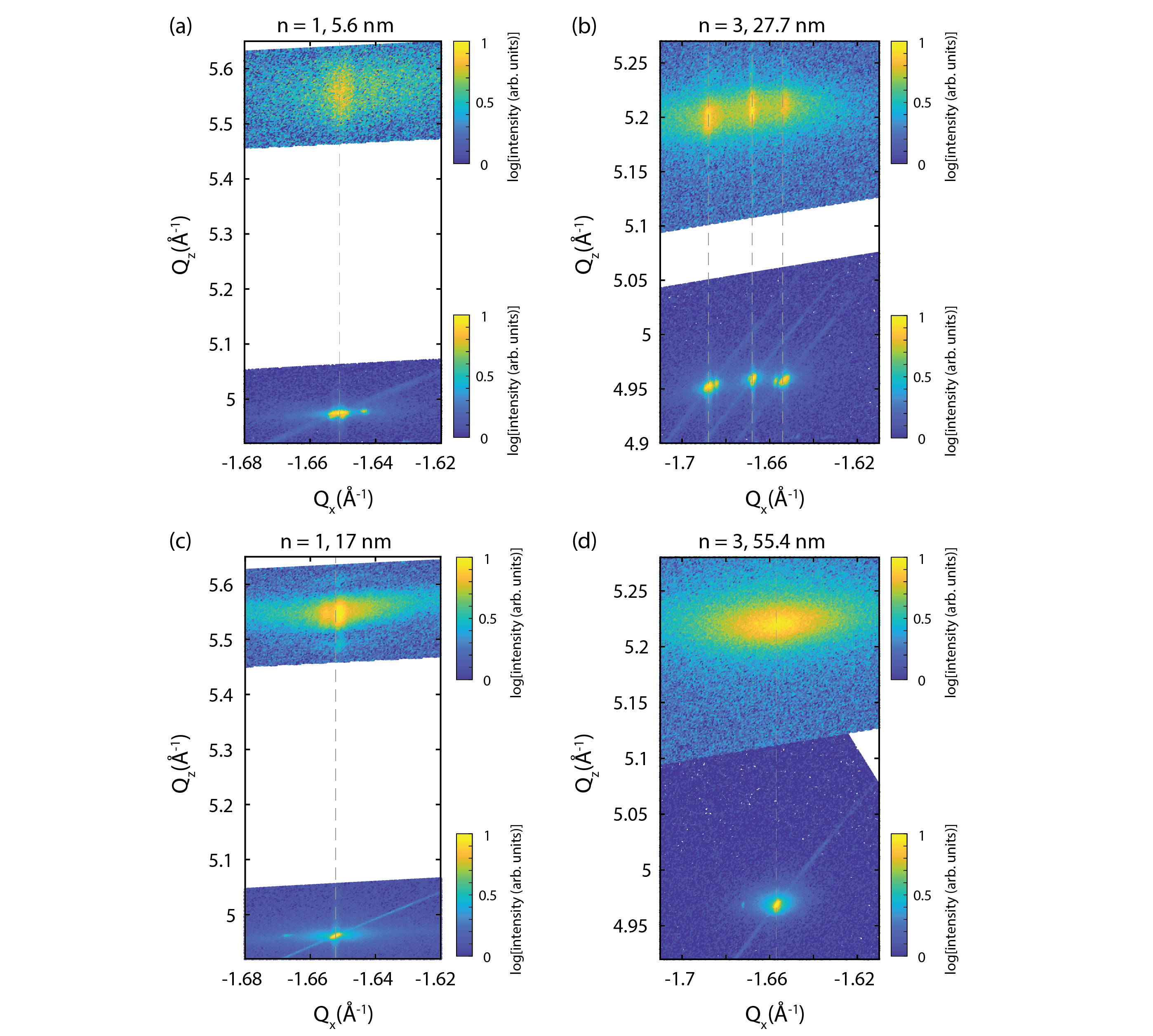}
    \caption{Reciprocal space maps of the Nd$_2$NiO$_4$ ($n = 1$) (a,c) and Nd$_4$Ni$_3$O$_{10}$ ($n = 3$) (b,d) for thinner and thicker films. Dashed lines are centered around the substrate LaAlO$_3$ (103) peaks and serve as guides to the eye.  For clarity, the color scales have been scaled independently for the substrate and film peaks.}
    \label{fig:rsm2}
\end{figure}

\newpage

\section*{Variation in film properties}

\begin{figure}[!h]
    \centering
    \includegraphics[width = \columnwidth]{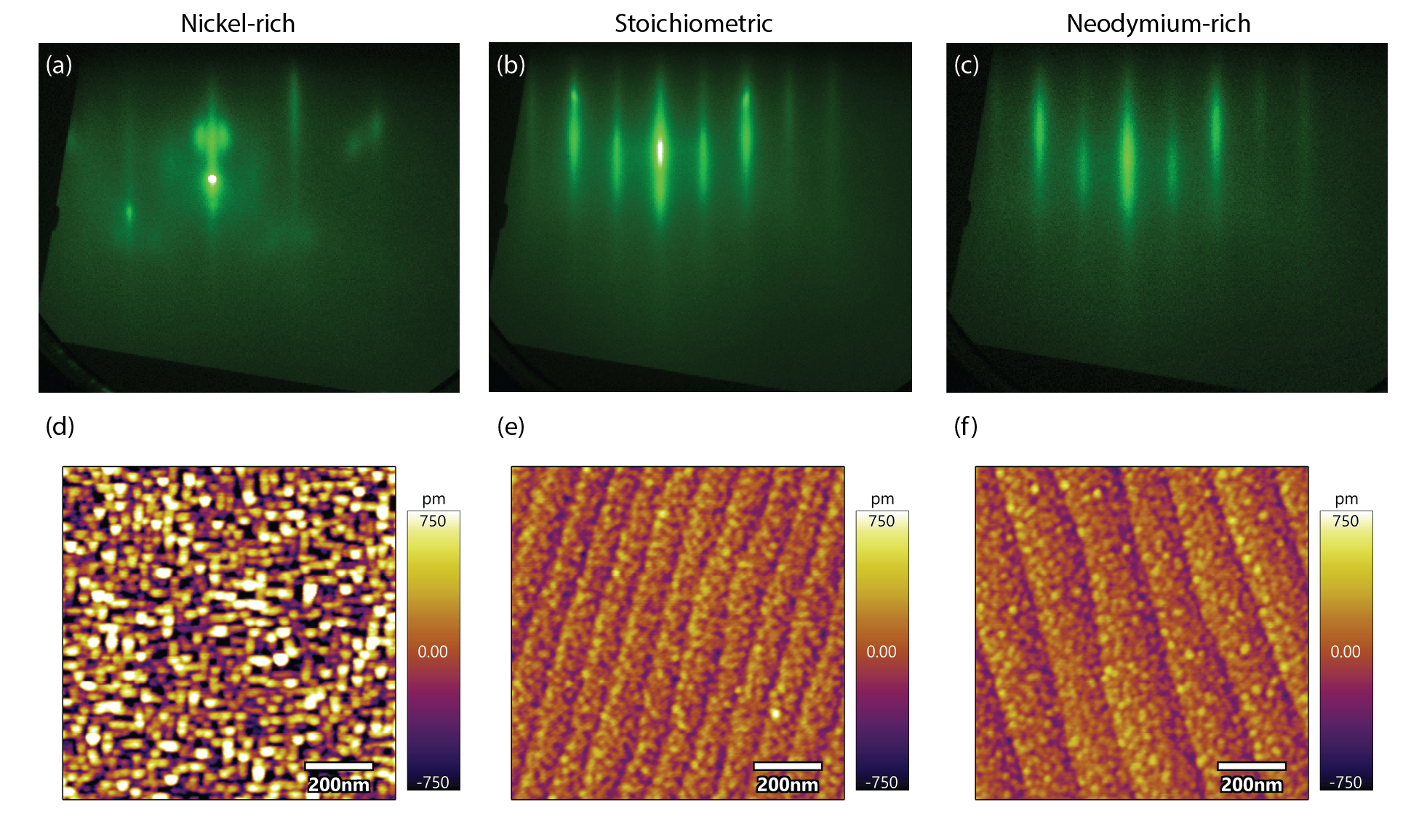}
    \caption{Reflection high-energy electron diffraction (RHEED) and atomic force microscope (AFM) images of NdNiO$_3$ films with excess nickel (a,d), stoichiometric compositions (b,e), and excess neodymium (c,f).  Note that excess neodymium is not obviously manifest in surface sensitive characterization like RHEED and AFM. }
    \label{fig:rheed}
\end{figure}

It is well-known that NdNiO$_3$ thin films can accommodate a wide range of non-stoichiometric compositions with dramatic effects on the electrical transport properties, as thoroughly documented by Breckenfield \textit{et.\ al.\ } \cite{breckenfeld2014effects} and which we have described in the main text.  In Fig.\ \ref{fig:rheed}, we provide characterization of the surface of examples of our own non-stoichiometric NdNiO$_3$ thin films, which we use to calibrate the Ruddlesden-Popper compounds.  While excess nickel is manifest in the form of phase impurities in the reflection high-energy electron diffraction (RHEED) pattern and cubic NiO impurities in the atomic force microscope (AFM) image, excess neodymium is less distinct.

\begin{figure}[!h]
    \centering
    \includegraphics[width = \columnwidth]{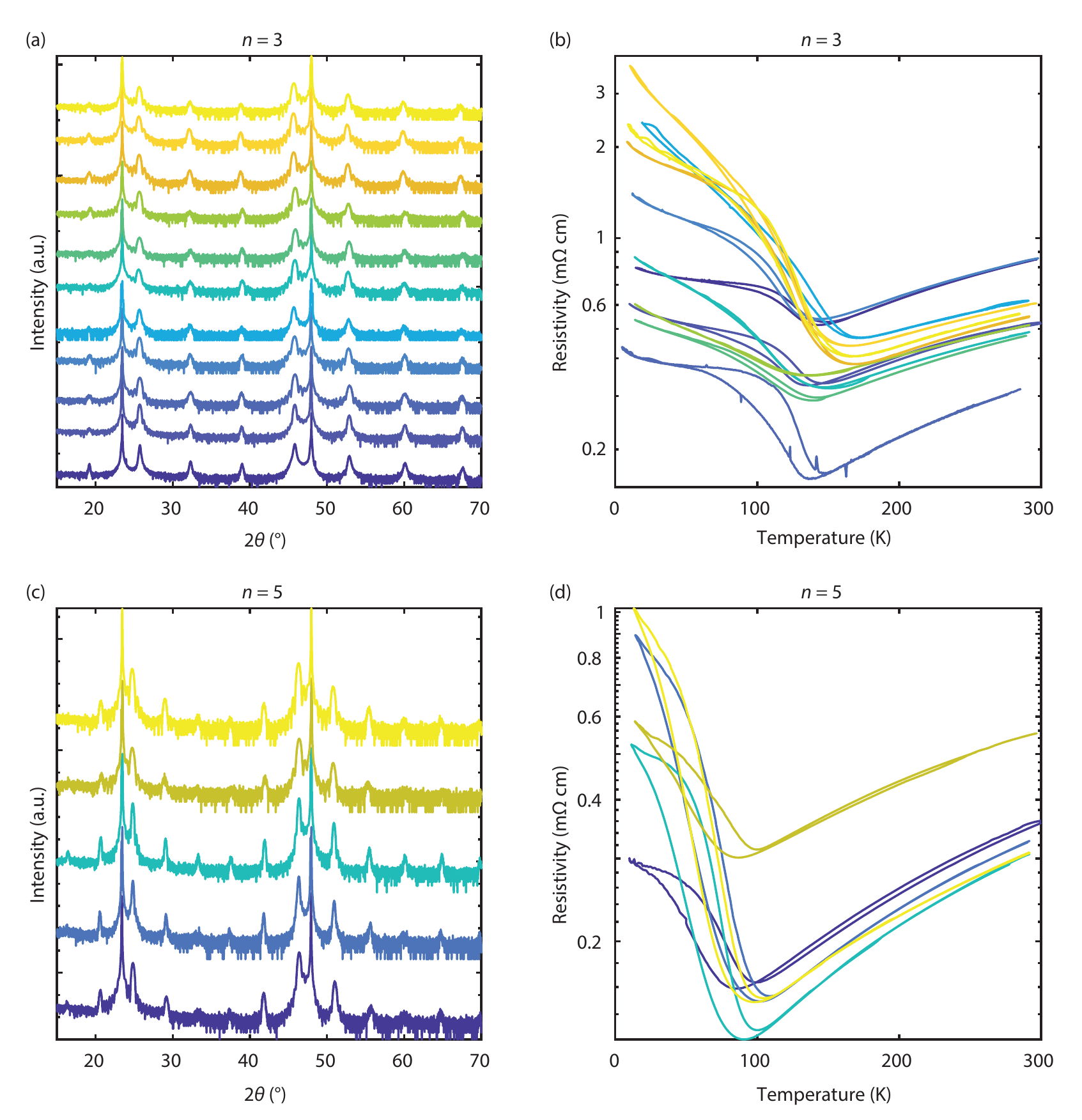}
    \caption{X-ray diffraction spectra (a,c) for various $n = 3$ and $n = 5$ thin films with the corresponding electrical resistivities (b,d).  The diffraction spectra are plotted in order of increasing out-of-plane lattice constant going from bottom to top, as determined by the method described in the previous section.}
    \label{fig:variablefilms}
\end{figure}

We note that Nd$_{n+1}$Ni$_n$O$_{3n+1}$ Ruddlesden-Popper compounds can also accommodate minor stoichiometric deviations, generally in the form of extra Ruddlesden-Popper layerings.  In Fig.\ \ref{fig:variablefilms}, we present x-ray diffraction spectra for many high quality specimens of $n = 3$ and $n = 5$ films.  These films are plotted in order of increasing lattice constant, determined following the procedure in the preceding section, from bottom to top. Despite slight structural variations, we consistently observe a metal-to-insulator transition in all of our Ruddlesden-Popper thin films synthesized on the substrate LaAlO$_3$, which provides the best lattice match to the bulk $n = 1, 3, \infty$ compounds ($< 1\%$).  While metal-to-insulator transitions could be expected in the higher-order $n = 5$ films, since $n = 5$ begins to approximate the perovskite NdNiO$_3$, metal-to-insulator transitions are surprisingly also observed in the lower-order $n = 3$ films, which typically exhibit metal-to-metal transitions when synthesized as bulk single crystals or powders.  The exact transition temperature $T_{\mathrm{MIT}}$ varies within a $\sim$ 30 K range film-to-film but is consistently higher for the $n = 3$ films ($T_{\mathrm{MIT}} \sim$ 130 K $-$ 150 K) than for the $n = 5$ films ($T_{\mathrm{MIT}}\sim 80$ K $-$ 110 K).  Ultimately, at the macroscopic level there does not appear to be any obvious correlation between lattice constant (or Nelson-Riley fitting error) and $T_{\mathrm{MIT}}$ or the width of the hysteresis.

\section*{Microstructural characterization with scanning transmission electron microscopy}

Here we present additional characterization of film microstructure and defects using scanning transmission electron microscopy (STEM) and a phase lock-in analysis on the high-angle annular dark-field (HAADF) images.

A phase lock-in analysis of HAADF-STEM images enables a quantitative assessment of the distribution of the local $n$ layerings in a film.  We perform our analysis on the (001) and (101) pseudocubic peaks, which both highlight the strong phase shifts that occur at Ruddlesden-Popper boundaries due to changes in the interplanar (fringe) spacing between consecutive Nd-O layers and between Nd-O and Ni-O layers. By taking the gradient of the local phase within the HAADF-STEM image, we can produce a map of local lattice strain along the $c$-axis of the pseudocubic crystal, which highlights the horizontal Ruddlesden-Popper rock salt layers. These rock salt layers are highlighted in analysis of the (001) pseudocubic peak due to the change in the interplanar spacing between consecutive Nd-O rocksalt layers ($\sim$ 2.7 \AA) compared to the pseudocubic c-axis lattice constant ($\sim$ 3.79 \AA), which is interpreted as an effective contraction of the lattice constant.  In analysis of the (101) pseudocubic peak, this boundary is also highlighted due to the similarly decreased spacing between local atoms in the [101] direction of the lattice. 

We chose these two different peaks to optimize the clean identification of these boundaries, which helps overcome the effects of mixed projection in the images caused by the electron beam encountering stacking faults as it passes through the sample.  This mixed projection through both neodymium- and nickel-site columns produces HAADF-STEM images with reduced contrast and an obscured view of the Ruddlesden-Popper boundaries.

Taking profiles of the vertical atomic planes, we extracted the distance between two Ruddlesden-Popper boundaries from the fringe strain map, from which the local $n$ phase can be measured. This analysis can be repeated across HAADF-STEM images to yield statistical counts of local $n$, as shown in Fig.\ 2. 

\begin{figure}[!h]
    \centering
    \includegraphics[width = \columnwidth]{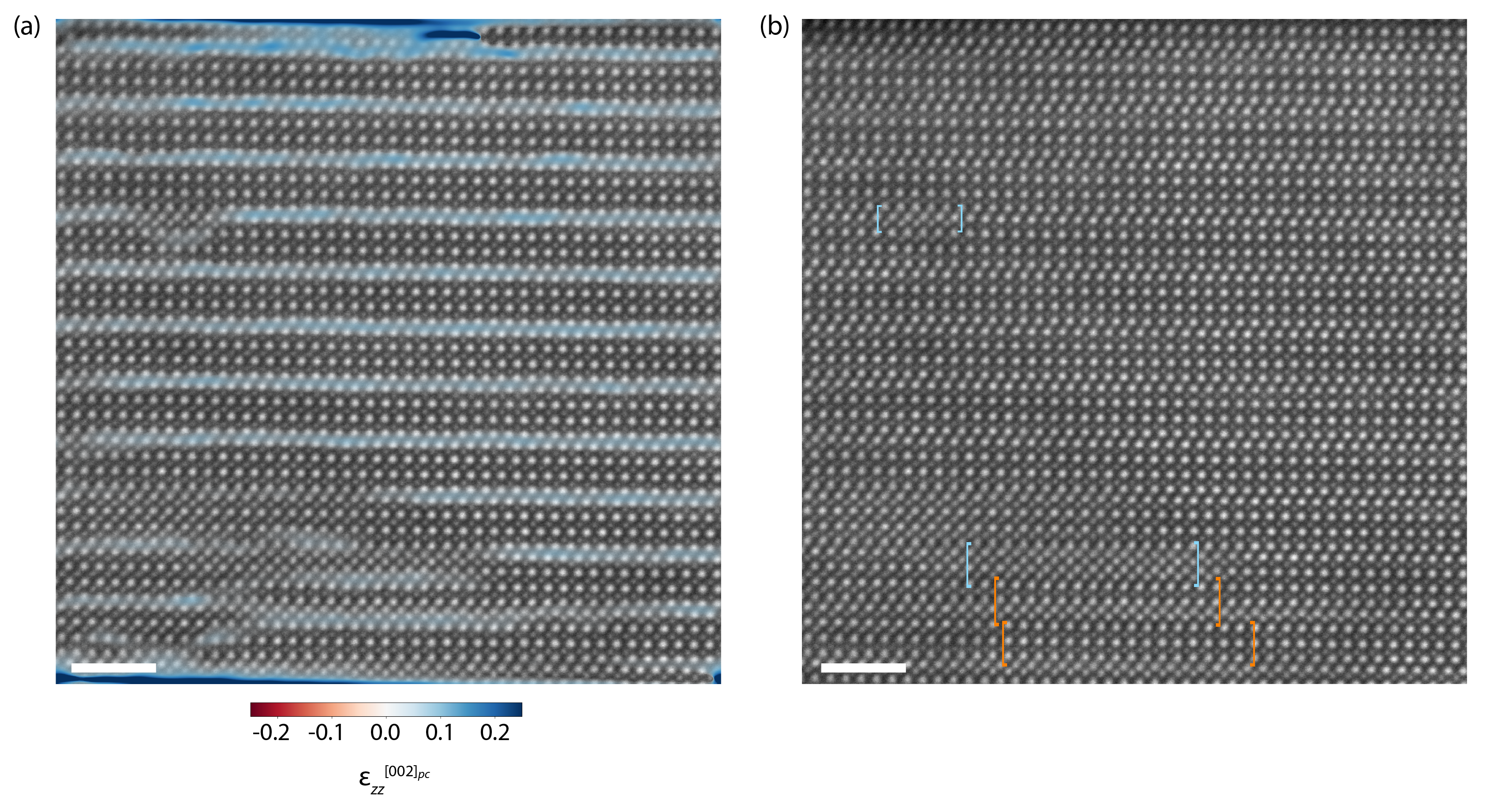}
    \caption{Examples of observed local defect structures in an $n = 3$ thin film via STEM imaging, with (a) and without (b) an overlaid strain map.  Discontinuous rock salt layers are clearly visible in the strain map produced by phase lock-in analysis. Regions of reduced atomic contrast, indicated by the blue brackets in (b), can frequently be found near the rock salt discontinuities.  These regions are indicative of half-unit-cell offsets, i.e by $a$/2 in the [100]$_{pc}$ direction, potentially due to stacking faults.  Any defect may lead to regions of locally varying $n$.  Orange brackets in (b) indicate regions that are better indexed as $n = 2$ than $n = 3$. Scale bars, 2 nm.}
    \label{fig:defect}
\end{figure}

We note that the SrTiO$_3$-based Ruddlesden-Popper compounds may possess more mobile or missing rock-salt layers that lead to large regions of high $n$ \cite{dawley2020defect,nie2014atomically,lee2014dynamic}.  Applying the phase lock-in analysis to the NdNiO$_3$ Ruddlesden-Popper compounds however, we do not frequently observe these missing rock salt layers in the largest field-of-view scanning transmission electron microscopy images.  The primary defects we observe in our films include discontinuous Nd-O rock salt layers that lead to locally varying $n$ for small $n$ ($\leq 6$) and half-unit-cell offsets in the [100]$_{pc}$ direction, as we document in Fig.\ \ref{fig:defect}.  Hence there do not seem to be large regions of NdNiO$_3$ that would substantially contribute to the metal-to-insulator transitions we see.

\begin{figure}[!h]
    \centering
    \includegraphics[width = \columnwidth]{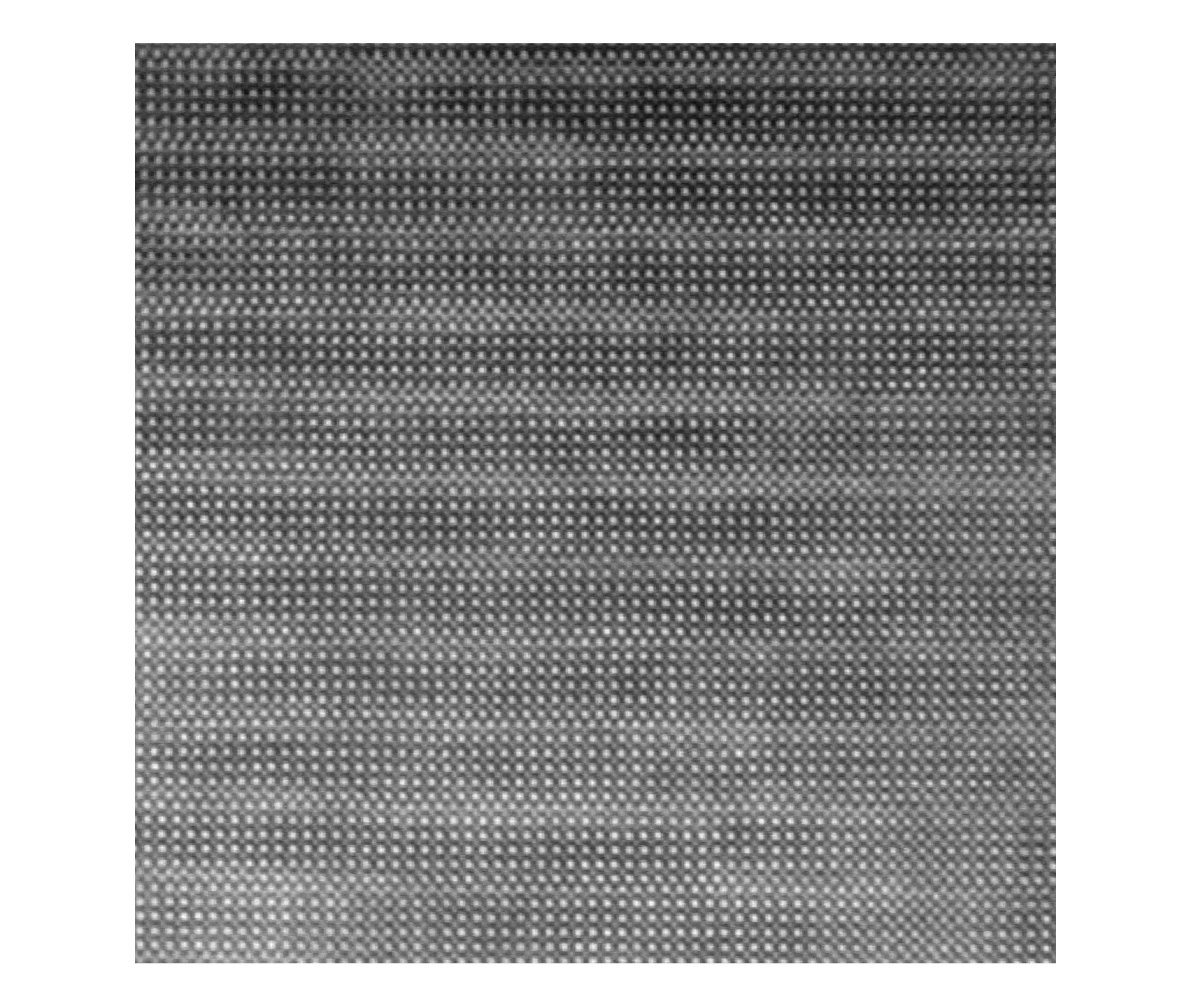}
    \caption{The raw HAADF-STEM image of the $n = 5$ compound used to generate the strain map in Fig.\ 2a of the main text.}
    \label{fig:rawn5}
\end{figure}

\newpage

\newpage
\begin{figure}[!h]
    \centering
    \includegraphics[width = \columnwidth]{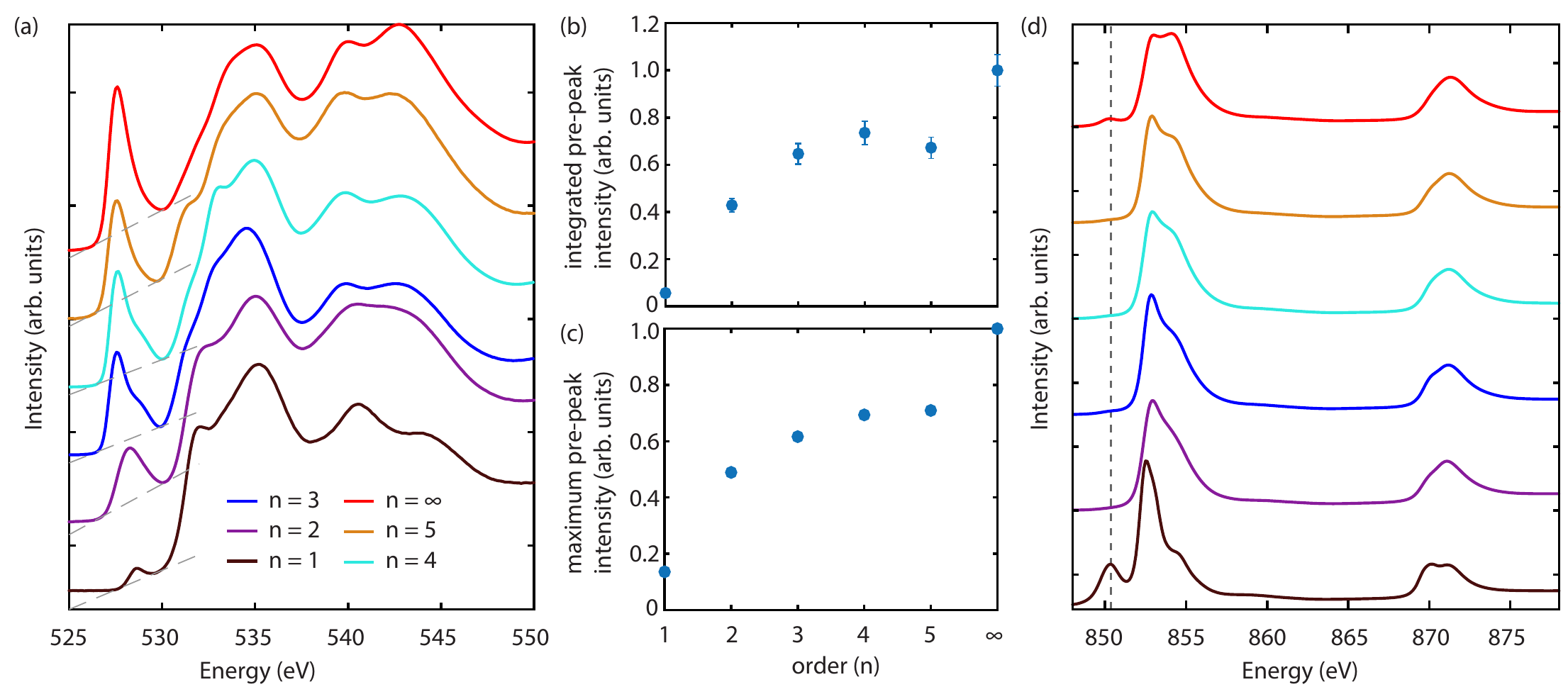}
    \caption{X-ray absorption spectra of the entire (a) O-$K$ edges.  We do not discuss the extended fine structure  of the O-$K$ edges in the main text due to potential contributions from the substrate background or trace surface NiO impurities.  However, the pre-peak region of the Nd$_{n+1}$Ni$_n$O$_{3n+1}$ does not overlap with pre-peak features from substrate or NiO contributions and represents pure Nd$_{n+1}$Ni$_n$O$_{3n+1}$.  The grey lines represent the linear backgrounds subtracted to quantify the integrated pre-peak intensities (b) and the maximum of pre-peak heights (c) following a process similar to that described in ref.\ \onlinecite{suntivich2014estimating}.  Errors in (b) are determined from errors in estimating the integration window.  (d)  XAS of the entire Ni-$L_{2,3}$ edges. The Ni-$L_3$ edges were excluded from the non-negative linear least squares fitting employed in the main text to avoid contributions from the La-$M_4$ edge of the LaAlO$_3$ substrate.  This signal is present in the Nd$_2$NiO$_4$ and NdNiO$_3$ films due to exposed substrate from film patterning.}
    \label{fig:xasfull}
\end{figure}

\clearpage
\newpage
\bibliography{bib.bib}